\documentclass[a4paper,12pt]{amsart}

\usepackage[left=1in, right=1in, top = 1.5in, bottom = 1.5in]{geometry}
\usepackage[utf8]{inputenc}
\usepackage[all]{xy}
\usepackage{amssymb}
\usepackage{amsmath}
\usepackage{csquotes}
\usepackage{subfig}
\usepackage{enumerate}
\usepackage{graphicx}
\usepackage{amsaddr}
\usepackage{color}
\usepackage{mwe}
\usepackage{subfig}
\usepackage{appendix}
\usepackage{dsfont}
\usepackage{natbib}
\usepackage{url}

\usepackage{soul}
\usepackage{xcolor}

\definecolor{ao}{rgb}{0.0, 0.5, 0.0}

\usepackage{tikz}
\usepackage{pgfplots}
\usetikzlibrary{patterns, intersections}
\usepgfplotslibrary{fillbetween}
\pgfplotsset{compat=1.18}

\usepackage{setspace}
\usepackage[linesnumbered,ruled,vlined]{algorithm2e}
\newtheorem{lemma}{Lemma}

\newtheorem{theorem}{Theorem}
\newtheorem{corollary}{Corollary}

\newtheorem{assumption}{Assumption}
\theoremstyle{definition}

\newcommand{\indep}{\rotatebox[origin=c]{90}{$\models$}}

\MakeOuterQuote{"}\EnableQuotes
\DeclareUnicodeCharacter{00A0}{ }

\title[]{Nonparametric tests of treatment effect homogeneity for policy-makers}

\author{Oliver Dukes$^{1}$, Mats J. Stensrud$^{2}$, Riccardo Brioschi$^{2}$,\\ Aaron Hudson$^{3}$} \address{ $^1$ Department of Mathematics, Statistics and Computer Science,\\ Ghent University, Belgium. \\
$^2$ Department of Mathematics, École Polytechnique Fédérale de Lausanne, Switzerland.\\
$^3$ Vaccine and Infectious Disease Division, Fred Hutchinson Cancer Center, USA.}

\begin{document}
\maketitle

\begin{abstract}
Recent work has focused on nonparametric estimation of conditional treatment effects, but inference has remained relatively unexplored. We propose a class of nonparametric tests for both quantitative and qualitative treatment effect heterogeneity. The tests can incorporate a variety of structured assumptions on the conditional average treatment effect, allow for both continuous and discrete covariates, and do not require sample splitting to obtain a tractable asymptotic null distribution. Furthermore, we show how the tests are tailored to detect alternatives where the population impact of adopting a personalized decision rule differs from using a rule that discards covariates. The proposal is thus relevant for guiding treatment policies. The utility of the proposal is borne out in simulation studies and a re-analysis of an AIDS clinical trial. \end{abstract}

\section{Introduction}

Many studies aim to investigate how treatment effects vary between groups of individuals. What we call effect heterogeneity is often referred to as an \textit{interaction} in the statistics literature, meaning that the treatment effect on a relevant outcome depends on certain patient characteristics.\footnote{Henceforth, we intentionally avoid using the term interaction, because the term has a different, interventional interpretation in the causal inference literature \citep{vanderweele2009distinction}.}
The existence of effect heterogeneity is a premise for the model of personalised medicine, where treatment decisions are made for specific sub-populations of patients. 

More specifically, \textit{quantitative heterogeneity} occurs when the effectiveness of the treatment varies by subgroup. Studying quantitative heterogeneity can reveal important differences in the effectiveness of treatment. However, to make decisions, it is often relevant to study whether treatment is beneficial for certain subgroups and harmful for others. Such \textit{qualitative heterogeneity}, also called qualitative effect modification, is of clinical interest when treatment decisions will be tailored to individual characteristics.  Qualitative heterogeneity also has the advantage that it does not depend on a given scale, whereas the absence of quantitative heterogeneity on one scale typically implies its presence on another. 

This paper concerns inference on both quantitative and qualitative heterogeneity in treatment effects.  These types of heterogeneity have been well studied when making comparisons between a small number of subgroups. One can infer quantitative heterogeneity by first obtaining estimators of the average treatment effect (ATE) in each subgroup, and subsequently using the estimators to construct a test of equality of the subgroup treatment effects. Similarly, one can test for qualitative heterogeneity using the likelihood ratio test of \citet{gail1985testing}, the \textit{range test} of \citet{piantadosi1993comparison} or other related approaches. The problem is more challenging when there are a large number or even infinitely many subgroups (e.g. with a continuous effect modifier). Estimating the subgroup effects becomes difficult due to the small sample size in each group, resulting in a potentially disastrous power loss. In principle, one could combine subgroups together (discretizing the continuous variable), or assume a parametric model for the conditional average treatment effect (CATE). However, it may be difficult to combine subgroups in a way that maximises power, and simple parametric models run the risk of mis-specification when the model is too simple (in turn also compromising power).

Although most focus has been given to estimation of heterogeneous effects \citep{kennedy2023towards}, some nonparametric tests for treatment effect heterogeneity have been described. A nonparametric approach is attractive as it may give power to detect a more flexible class of alternatives that could be missed by a more restrictive parametric strategy. For quantitative heterogeneity, \citet{crump2008nonparametric} proposed a Wald test based on series estimation of the CATE and \citet{ding2019decomposing} justified a related parametric approach solely under the randomization of treatment. \citet{chernozhukov2025fisher} and \citet{sanchez2023robust} provided extensions of these tests that incorporate machine learning. Some nonparametric tests of whether the CATE is non-negative (or non-positive) over the covariate support have also been described, which are closer to our own work. \citet{chang2015nonparametric} proposed a test based on an $L_1$-functional of a kernel smoothing estimator of the CATE, whilst \citet{hsu2017consistent} described a Kolmogorov-Smirnov test using a hypercube kernel. \citet{shi2019sparse} develop two tests based on the implication of the null hypothesis of non-negativity that the average response under the optimal dynamic treatment rule equals the average response under a `treat-everyone' rule. They use a plug-in estimator of the optimal rule (e.g. treat individuals if the estimated CATE exceeds zero) obtained from a separate sample than the one used to construct the test statistic. Sample-splitting enables \citet{shi2019sparse} to show that their test statistic converges to a limiting normal distribution under the null. Unlike splitting in semiparametric inference to weaken assumptions on nuisance parameters estimation \citep{chernozhukov2018double}, it is not clear how power can be recovered in this context via cross-fitting (swapping test and training samples then averaging results), whilst preserving a tractable limiting null distribution. This is due to the complex dependency of test statistics across folds.

In this paper, we propose a class of nonparametric tests for both quantitative and qualitative heterogeneity. Compared to existing proposals, our tests can incorporate a variety of structured assumptions on the CATE and retain validity even if these assumptions fail. Moreover, they extend to moderate-dimensional covariates, and they do not require sample splitting. Loosely, our tests have non-trivial power when implementing an individualized decision rule within a class of choice would lead to a different outcome (at the population level) than ignoring covariates. They are therefore useful in settings where potential heterogeneity might lead to policy changes. Our work builds upon that of \citet{hsu2017consistent}, although our null hypotheses differ and we consider generalized implementations beyond the hypercube kernel, drawing instead on \textit{empirical welfare maximization} \citep{kitagawa2018should,athey2021policy}. Like \citet{shi2019sparse}, we connect the testing problem with inference on the optimal value; however, we also consider quantitative heterogeneity, do not use a plug-in estimator of the optimal rule and our test statistics have a tractable asymptotic null distribution without requiring sample splitting. Using the full sample to construct the test statistic is expected to confer benefits in terms of power. Our theory of local asymptotics is also distinct from previous work. The inferential strategy we take is related to that of \citet{li2024estimation}, although they consider estimation of performance metrics for policy learning, rather than testing for effect heterogeneity.

\section{Preliminaries}\label{sec:prelim}

\subsection{Notation and review}

Consider data of the form $Z = (X, A, Y)$, where $X \in \mathbb{R}^p$ is a covariate vector, $A$ is a binary treatment, and $Y$ is a real-valued outcome. Then $Z_1,...,Z_n$ represent $n$ i.i.d. draws from a data law $P_0$, which belongs to a nonparametric model $\mathcal{M}$. For $s \subseteq \{1, \ldots, p\}$, let $X_s$ be the subvector of $X$ containing elements with indices belonging to $s$. Let $Y(a)$ denote the counterfactual outcome under treatment $A = a$.  We let $E_0$ refer to an expectation taken under the law $P_0$, whereas $E_P$ is taken with respect to an arbitrary law $P$ in $\mathcal{M}$. Similarly, we denote the probability of an event occurring under an arbitrary law by $Pr_P$ and let $Pr_0$ be the probability taken under $P_0$.
Let $\tau_{P} := E_P\{Y(1) - Y(0)\}$ denote the ATE, and let $\tau_{P,s}(x_s):= E_P\{Y(1) - Y(0)| X_s = x_s\}$ denote the CATE, with $\tau_{0,s}$ and $\tau_{0}$ representing their evaluations under $P_0$. Our set-up allows for testing for heterogeneity with respect to a subvector of $X$; we assume $X_s$ is selected in advance, based on the research question.

Before describing the inference problem, we first review existing results on identification of the CATE in randomized and observational studies. Let $\mu_P(a,x) :=  E_P(Y|A = a, X = x)$ denote the conditional mean of the outcome given the treatment and covariates. We also let $\pi_P(a|x) := Pr_P(A = a| X=x)$ denote the treatment assignment mechanism or \textit{propensity score}; $\mu_0(a,x)$ and $\pi_0(a|x)$ refer to these quantities evaluated at the true law $P_0$. We will make the following assumptions:
\begin{assumption}\label{consistency}
(Consistency) If $A=a$, then $Y=Y(a)$.
\end{assumption}
\begin{assumption}\label{positivity}(Positivity) If $f_X(x)>0$ then $\pi_0(a|x)>0$ for $a\in \{0,1\}$, where $f_X(x)$ is the density of $X$.
\end{assumption}
\begin{assumption}\label{cond_ex}
(Conditional exchangeability) $Y(a)\indep A|X$ for $a=0,1$. 
\end{assumption}
Under the above assumptions, the ATE and CATE, respectively, are identified as 
\begin{align*}
\tau_0 = E_0\left\{\mu_0(1,X) - \mu_0(0,X) \right\}, 
\quad
\tau_{0,s}(x_s)=E_0\left\{\mu_0(1, X) - \mu_0(0, X)|X_s=x_s\right\}.
\end{align*}

\subsection{Estimands for quantitative effect heterogeneity}

As argued in the introduction, testing for heterogeneity is more challenging when potential effect modifiers are continuous and/or moderate-dimensional. In order to motivate our test, we will propose estimands that summarise effects at different levels of the covariates in a data-driven way. These estimands are generic in the sense that they apply for covariates that are discrete or continuous, scalar or multivariate.

\begin{figure}
    \centering
    \includegraphics[width=0.9\linewidth]{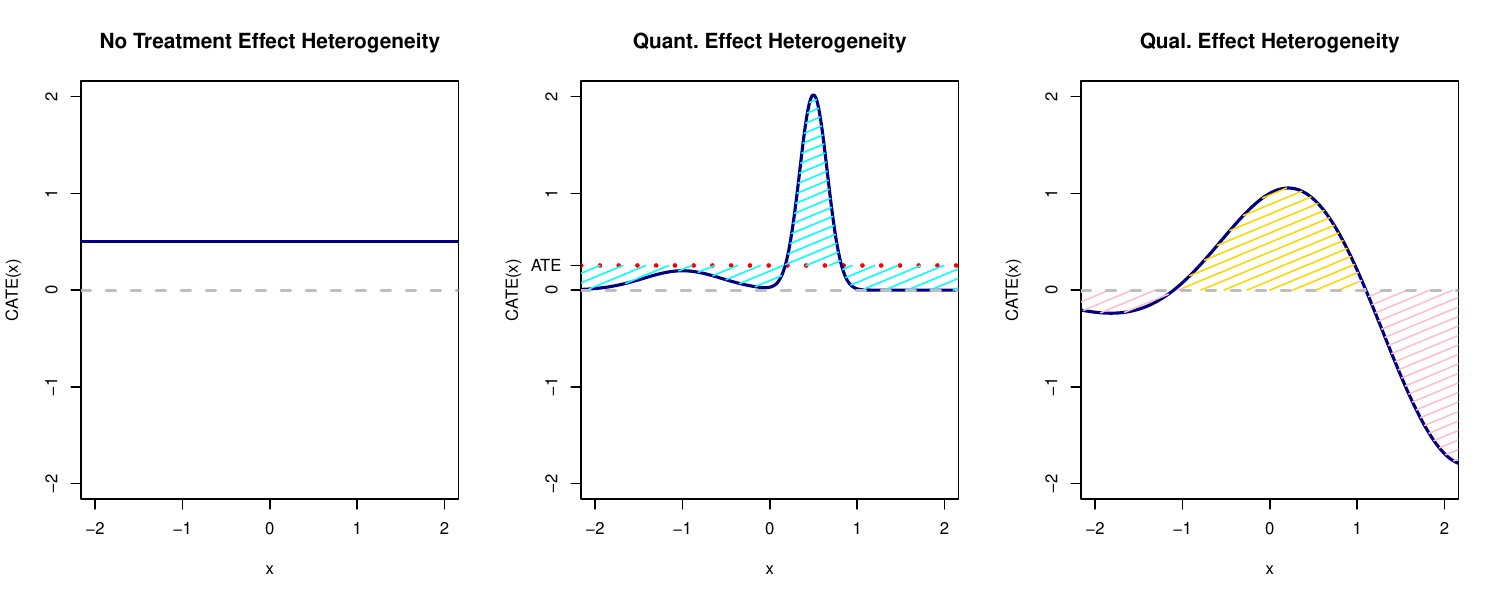}
 \caption{Illustration of effect heterogeneity.}
    \label{fig:interaction}
\end{figure}

We begin by describing an estimand for quantitative effect heterogeneity.
Consider
\begin{align*}
    \theta_{P,\tau_P}^+ &:= E_P[\{\tau_{P,s}(X_s) - \tau_P\}\mathds{1}(\tau_{P,s}(X_s) \geq \tau_P)],
    \\
    \theta_{P,\tau_P}^- &:= E_P[\{\tau_{P,s}(X_s) - \tau_P\}\mathds{1}(\tau_{P,s}(X_s) \leq \tau_P)],
\end{align*}
where we will use the shorthand notation $\theta^+_{0,\tau_{0}} := \theta^+_{P_0,\tau_{P_0}}$ and $\theta^-_{0,\tau_{0}} := \theta^-_{P_0,\tau_{P_0}}$.
To give some intuition, suppose that $X_s$ is a scalar continuous covariate that is uniformly distributed on a fixed interval; see Figure \ref{fig:interaction} (center) for an example plot of the CATE against $X_s$.
Then $\theta_{0,\tau_0}^+$ ($\theta_{0,\tau_0}^-$) represents the area above (below) the mean-centered conditional treatment effect curve after appropriate scaling, measuring the extent to which the treatment performs better (worse) than average.
Moreover, it can easily be seen that
\begin{align*}
    \theta_{0,\tau_0}^+-\theta_{0,\tau_0}^- = E_0\{|\tau_{0,s}(X_s) - \tau_0|\},
\end{align*}
giving us a representation of the probability-weighted $L_1$-distance of the CATE curve from the mean. Given this intuition, we believe that this is often easily interpretable as a summary of heterogeneity relative to contrasts based on other distances (e.g. $L_2$-distance).

We can assess additive heterogeneity by determining whether the area above or below the centered CATE curve is zero.
Formally, our goal is to test the null hypothesis
\begin{align*}
    H^{\mathrm{I}}_0: \theta_{0,\tau_0}^+ - \theta_{0,\tau_0}^- = 0,
\end{align*}
against its complement. We note that this is a test specifically of \textit{additive} effect heterogeneity. Heterogeneity on the additive scale is arguably the most relevant in public health decision-making \citep{rothman1980concepts}, since policy-makers are often interested in the absolute causal effects of interventions and how they may differ between subgroups. Extending our framework to other scales is a topic for future work. 

\subsection{Estimands for qualitative effect heterogeneity} \label{sec:qual-estimands}
 
We define the estimands
\begin{align*}
    &\theta^{+}_{P,\delta} := E_P\left[\left\{\tau_{P,s}(X_s) - \delta\right\} \mathds{1}(\tau_{P,s}(X_s) \geq \delta )\right],
    \\
    &\theta^{-}_{P,\delta} := E_P\left[\left\{\tau_{P,s}(X_s) - \delta \right\} \mathds{1}(\tau_{P,s}(X_s) \leq \delta )\right],
\end{align*} 
for $\delta \in \mathbb{R}$, and use the shorthand notation $\theta^+_{0,\delta} := \theta^+_{P_0,\delta}$ and $\theta^-_{0, \delta} := \theta^{-}_{P_0, \delta}$.
Here, $\theta^+_{0,\delta}$ is positive whenever the CATE exceeds $\delta$ on a set with measure greater than zero, and $\theta^+_{0,\delta}$ is zero otherwise. Similarly, $\theta^-_{0,\delta}$ is negative whenever the conditional treatment effect is below $\delta$ on a set with measure greater than zero. In Figure (right), we set $\delta=0$; the harmful (beneficial) effect of treatment is represented by the area under the curve below (above) the $x$-axis. 

We emphasize that $\delta$ is not a tuning parameter. Rather, different choices correspond with different scientific hypotheses that can be of interest in a given setting. For instance, $\delta=\tau_0$ coincides with a test of quantitative heterogeneity; $\delta=0$ coincides with a test of qualitative heterogeneity. In certain settings, one might choose $\delta$ to correspond to a `clinically important' benefit, which must be specified in advance. For example, COVID 19 vaccines were expected to have at least 50\% efficacy in placebo-controlled trials \citep{krause2020covid}. Hence one might investigate the existence of subgroups with efficacy above/below 50\%.

If both $\theta^+_{0,\delta}$ and $\theta^-_{0,\delta}$ are non-zero, there is evidence for a qualitative difference in treatment effects. In particular, with $\delta = 0$, this corresponds to qualitative heterogeneity in the sense of \citet{gail1985testing}, meaning that the treatment is beneficial for a subset of the population and harmful to another subset. Qualitative heterogeneity is implied by quantitative heterogeneity, and the latter will usually be easier to detect. On the other hand, qualitative heterogeneity is scale independent and may have clearer policy ramifications. Indeed, qualitative heterogeneity is more relevant when there is interest in tailoring treatment strategies to avoid potential harm. When qualitative heterogeneity could be expected, $\theta^+_{0,\delta}$ and $\theta^-_{0,\delta}$ might also be of independent interest as summaries of heterogeneity.

Formally, our inferential goal is to construct a test of the null hypothesis
\begin{align*}
    H^{\mathrm{II}}_0: \theta^+_{0,\delta} = 0 \textbf{ or } \theta^-_{0,\delta} = 0
    \end{align*}
against its complement. This is an example of a \textit{composite null hypothesis}, meaning there are a range of values that $\theta^+_{0,\delta}$ and $\theta^+_{0,\delta}$ can take which are compatible with the null. If we are interested only in whether the treatment effect falls below $\delta$ for some population members, then our null would become $\theta^-_{0,\delta}=0$. 

\subsection{Identification}
At an arbitrary law $P$ in $\mathcal{M}$, we define 
\begin{align}
g_{P,\delta}: w \mapsto \mathds{1}(E_{P}\{\mu_P(1,X) - \mu_P(0,X)|X_s = w\} \geq \delta),
\label{opt-rule}
\end{align}
where $g_{0, \delta} := g_{P_0, \delta}$ under the true law. Under Assumptions 1-3, it follows from the law of total expectation that $\theta^+_{0,\delta}$ and $\theta^-_{0,\delta}$ can be expressed as
\begin{align*}
    &\theta^+_{0,\delta} = E_0\left[\left\{\mu_0(1, X) - \mu_0(0, X)-\delta\right\}g_{0,\delta}(X_s)\right] 
    \\
    &\theta^-_{0,\delta} = E_0\left[\left\{\mu_0(1, X) - \mu_0(0, X) -\delta\right\}\left\{1 - g_{0,\delta}(X_s) \right\}\right].
\end{align*}
 We now have identification functionals for $\theta^+_{0,\delta}$ and $\theta^-_{0,\delta}$ which can be estimated and used to construct a test based on the observed data. 

\section{Methodology}\label{sec:method}

\subsection{Strategy for inference}

In what follows, we will develop nonparametric inference for $\theta_{0,\tau_0}^+$, $\theta_{0,\tau_0}^-$, $\theta^+_{0,\delta}$ and $\theta^-_{0,\delta}$. The construction of estimators and hypothesis tests for smooth functionals  of the data-generating mechanism under a nonparametric model is now well-understood \citep{kennedy2024semiparametric}. However, our setting poses additional challenges. Firstly, 
 $\theta_{P,\tau_P}^+$, $\theta_{P,\tau_P}^-$, $\theta^+_{P,\delta}$ and $\theta^-_{P,\delta}$ involve indicators, which are non-differentiable functions, and hence are non-smooth. 
Secondly, $\theta_{P,\tau_P}^+$ and $\theta^+_{P,\delta}$ are non-negative, $\theta_{P,\tau_P}^-$ and  $\theta^-_{P,\delta}$ are non-positive and all equal zero under the null. Moreover, the efficient influence function for each parameter vanishes under the null hypothesis.
As a result, estimators based on the sample average of the efficient influence function will not attain characterizable limiting distributions under the null. 
Standard testing procedures may then fail to protect the type I error rate. 

Let $f: \mathbb{R}^{|s|} \to [0,1]$ be a fixed function, let $\theta^+_{P,\delta}(f)$ and $\theta^-_{P,\delta}(f)$ be defined by
\begin{align*}    \theta^+_{P,\delta}(f) &:= E_P[\{\mu_P(1, X) - \mu_P(0, X) - \delta \}f(X_s)] 
    \\
    \theta^-_{P,\delta}(f) &:= E_P[\{\mu_P(1, X) - \mu_P(0, X) - \delta \}\{1 - f(X_s) \}],
\end{align*}
and let $\theta^+_{0,\delta}(f):=\theta^+_{P_0,\delta}(f)$ and $\theta^-_{0,\delta}(f) := \theta^-_{P_0, \delta}(f)$. Note firstly that in general, $\theta^+_{0,\delta}(f)$ is not constrained to be non-negative, and $\theta^-_{0,\delta}(f)$ is not constrained to be non-positive.
Furthermore, we observe that for any $f$ and any $
\delta$, $\theta^+_{0,\delta}(f) \leq \theta^+_{0,\delta}$, $\theta^-_{0,\delta}(f) \geq \theta^-_{0,\delta}$
with equality when $f=g_{0,\delta}$ as defined in \eqref{opt-rule}. Therefore, $g_{0,\delta}$ is both the maximizer of $\theta^+_{0,\delta}(f)$ and the minimizer of $\theta^-_{0,\delta}(f)$; at that choice, $\theta^+_{0,\delta}(f)$ equals the original target $\theta^+_{0,\delta}$. Thus, if there exist functions $f_1$ and $f_2$ for which $\theta^+_{0,\delta}(f_1) > 0$, and $\theta^-_{0,\delta}(f_2) < 0$, we have sufficient evidence to reject the null hypothesis of no \textit{qualitative} effect heterogeneity $H^{\mathrm{II}}_0$. In addition, with $
\delta = \tau_0$, we have the relation
$\left|\theta^{+}_{0,\tau_0}(f) - \theta^{-}_{0,\tau_0}(f)\right| \leq \theta^+_{0,\tau_0} -\theta^{-}_{0,\tau_0}$, with equality when $f = g_{0,\tau_0}$.
Therefore, to reject the null of no \textit{quantitative} effect heterogeneity $H_0^{\mathrm{I}}$, it is sufficient to show that there exists $f_\tau$ such that $|\theta^{+}_{0,\tau_0}(f_\tau) - \theta^{-}_{0,\tau_0}(f_\tau)| > 0$.

Now, let $\mathcal{F}$ be a class of functions from $\mathbb{R}^{|s|} \to [0,1]$. Then for any choice of $f\in \mathcal{F}$, we can bound the original target using the supremum over $\mathcal{F}$ of $\theta_{0,\delta}^+(f)$ and the infimum over $\mathcal{F}$ of $\theta^-_{0,\delta}(f)$:
\begin{align}
    \sup_{f \in \mathcal{F}} \theta^+_{0,\delta}(f) \leq \theta^+_{0,\delta},\quad  \inf_{f \in \mathcal{F}} \theta^-_{0,\delta}(f) \geq \theta^-_{0,\delta},
    \label{qual-ineq}
\end{align}
with equality when $\mathcal{F}$ contains $g_{0,\delta}$. Therefore if $\mathcal{F}$ contains $f_1$ and $f_2$ such that $\theta^+_{0,\delta}(f_1) > 0$ and $\theta^{-}_{0,\delta}(f_2)< 0$, there is evidence of both a positive and negative effect.
Similarly, with $\delta=\tau_0$, the supremum over $\mathcal{F}$ of $|\theta^+_{0,\tau_0}(f) - \theta^-_{0,\tau_0}(f)|$ serves as a lower bound for $\theta_{0,\tau_0}^+ - \theta_{0,\tau_0}^-$. That is,
\begin{align}
    \sup_{f \in \mathcal{F}} \left|\theta^{+}_{0,\tau_0}(f) - \theta^{-}_{0,\tau_0}(f)\right| \leq \theta_{0,\tau_0}^+ - \theta_{0,\tau_0}^-,
    \label{quant-ineq}
\end{align}
with equality when $
\mathcal{F}$ contains $g_{0,\tau_0}$.

The above suggests that the following approaches may be used to construct tests for effect heterogeneity. To test for quantitative heterogeneity, we assess the hypothesis
\begin{align*}
    H^{\mathrm{I},*}_0: \sup_{f \in \mathcal{F}} |\theta^{+}_{0,\tau_0}(f) - \theta^{-}_{0,\tau_0}(f)| = 0
\end{align*}
against its complement. This can alternatively be represented as $H^{\mathrm{I},*}_0: \sup_{f \in \mathcal{F}} |2\theta^{+}_{0,\tau_0}(f)| = 0$ since $\theta^+_{0,\tau_0}(f^*)=-\theta^-_{0,\tau_0}(f^*)$; indeed, we later utilize this representation when implementing the test. To test for qualitative heterogeneity, we perform one-sided tests of the hypothesis that the supremum (infimum) exceeds (falls below) zero. Specifically, we test the hypothesis
\begin{align*}
    H_0^{\mathrm{II},*}: \sup_{f \in \mathcal{F}} \theta^+_{0,\delta}(f) \leq 0 \textbf{ or } \inf_{f \in \mathcal{F}} \theta^-_{0,\delta}(f) \geq 0
\end{align*}
against its complement. Although for simple choices of $\mathcal{F}$, one might first estimate $\theta^+_{0,\delta}(f)$ and $\theta^-_{0,\delta}(f)$ for every $f\in \mathcal{F}$, this may not be feasible when  $\mathcal{F}$ is a large class. As described below, for certain $\mathcal{F}$ we will target the supremum/infimum directly.
 
It is easily seen that if $H^{\mathrm{I}}_0$ holds, then $H^{\mathrm{I},*}_0$ must hold as well. There are analogous relations between $H^{\mathrm{II}}_0$ and $H^{\mathrm{II},*}_0$. This implies that our test is valid for any choice of $\mathcal{F}$, in the sense that type I error should be asymptotically controlled, even when $\mathcal{F}$ is misspecified. 

Conversely, $H^{\mathrm{I},*}_0$ may still hold even when $H^{\mathrm{I}}_0$ does not if $\mathcal{F}$ does not contain $g_{0,\delta}$ and $1 - g_{0,\delta}$. Hence the choice of $\mathcal{F}$ will inevitably affect the power of the test. To give some intuition, consider a function $f^*$ satisfying both $\theta^+_{0,\delta}(f^*) \leq \theta^+_{0,\delta}$ and $\theta^-_{0,\delta}(f^*) \geq \theta^-_{0,\delta}.$
Set $\delta=0$ and define the (potentially stochastic) rule $g_{f^*}$ as the one that assigns treatment with probability $f^*(x_s)$. Then $\theta^{\text{min}}_{0,0}(f^*) = \min \{\theta^+_{0,0}(f^*),-\theta^-_{0,0}(f^*)\}$ is the expected benefit of implementing $g_{f^*}$ compared to the best of two static rules: 
\begin{align*}
\theta^{\text{min}}_{0,0}(f^*)  =    E_0\{Y(g_{f^*})\}- \max [ E_0\{Y(0)\}, E_0\{Y(1)\}]   
\end{align*}
as also noted by \citet{shi2019sparse}. For quantitative heterogeneity, one can show that 
\begin{align*}
\theta^+_{0,\tau_0}(f^*)=E_0\{Y(g_{f^*})\}-E_0\{Y(1)\}E_0\{f^*(X_s)\}-E_0\{Y(0)\}E_0\{1-f^*(X_s)\},
\end{align*}
which expresses the difference between the dynamic rule and a rule which would assign treatment with probability $E_0\{f^*(X_s)\}$, ignoring an individual's covariates. We derive both of the above equalities in the supplementary material Appendix \ref{appendix:policy}. Hence our tests have high power when a dynamic rule in our class -- determined by our test statistic -- gives substantially different outcomes in the population compared to any deterministic static rule, or a random static rule that ignores covariates. We investigate the impact of misspecifying $\mathcal{F}$ in our simulations.

\subsection{Estimation at any $f\in \mathcal{F}$}
To construct tests for effect heterogeneity, we need to be able to estimate $\theta^+_{0,\delta}(f)$, $\theta^-_{0,\delta}(f)$, $\theta^+_{0,\tau_0}(f)$ and  $\theta^-_{0,\tau_0}(f)$ for any $f\in \mathcal{F}$. One could devise plug-in estimators based on estimators of $\mu_0(a,x)$. However, such approaches generally inherit plug-in bias from the nuisance function estimators. To facilitate the use of data-adaptive estimators of the nuisance parameters, we will base our inferences on the \textit{efficient influence functions} of our target parameters. Recall that the original estimands fail to be pathwise differentiable under the null hypothesis. In contrast, this is not the case for the parameters indexed by a fixed $f$. 
\begin{lemma}\label{lemma:eif}
(The efficient influence function) Consider $f$ as fixed, and define the transformation 
\begin{align*}
    \psi_P:(x,a,y)  \mapsto \mu_P(1, x) - \mu_P(0, x) + \left(2a - 1 \right)\left\{ \frac{y - \mu_P(a,x)}{\pi_P(a|x)}\right\}.
\end{align*}
For any $f\in \mathcal{F}$ and for any fixed and known $\delta$, $\theta^+_{P,\delta}(f)$ and $\theta^-_{P,\delta}(f)$ are pathwise differentiable in a nonparametric model, and their efficient influence functions are given respectively by 
\begin{align*}
\varphi^+_{P,\delta}(Z;f)&:=\{\psi_P(Z)-\delta\}f(X_s)-\theta^+_{P,\delta}(f)\\ \varphi^-_{P,\delta}(Z;f)&:=\{\psi_P(Z)-\delta\}\{1-f(X_s)\}-\theta^-_{P,\delta}(f).
\end{align*}
Moreover, $\theta^{+}_{P,\tau_P}(f)$
and $\theta^{-}_{P,\tau_P}(f)$ are also pathwise differentiable in a nonparametric model, and their efficient influence functions are given respectively by
\begin{align*}
    \varphi^+_{P,\tau_P}(Z;f)&:=\{\psi_P(Z) - \tau_P\}\left[f(X_s) - E_P\{f(X_s)\}\right]-\theta^+_{P,\tau_P}(f)\\ \varphi^-_{P,\tau_P}(Z;f)&:=-\{\psi_P(Z)-\tau_P\}\left[f(X_s) -E_P\{f(X_s)\}\right] -\theta^-_{P,\tau_P}(f).
\end{align*}
\end{lemma}
A proof of this result, along with all others, is given in supplementary material Appendix \ref{appendix:proofs}. Lemma \ref{lemma:eif} is not readily useful for estimation because $\varphi^+_{P,\delta}(Z;f)$, $\varphi^-_{P,\delta}(Z;f)$, $\varphi^+_{P,\tau_P}(Z;f)$ and $\varphi^-_{P,\tau_P}(Z;f)$ depend on nuisance parameters that are in general unknown. Suppose then we have available estimators $\mu_n(a,x)$ and $\pi_n(a|x)$ for $\mu_0(a,x)$ and $\pi_0(a|x)$, and let $\psi_n$ be 
\begin{align*}
    \psi_n:(x,a,y)  \mapsto \mu_n(1, x) - \mu_n(0, x) + \left(2a - 1 \right)\left\{ \frac{y - \mu_n(a,x)}{\pi_n(a|x)}\right\}.
\end{align*}
We can construct one-step estimators for $\theta^+_{0,\delta}(f)$ and $\theta^-_{0,\delta}(f)$ as
\begin{align*}
    &\theta^{+}_{n,\delta}(f) := \frac{1}{n}\sum_{i=1}^n \left\{ \psi_n\left(Z_i\right) - \delta \right\} f(X_{s,i}),
    \\
    &\theta^{-}_{n,\delta}(f) := \frac{1}{n}\sum_{i=1}^n \left\{ \psi_n\left(Z_i\right) - \delta \right\} \left\{1 - f(X_{s,i}) \right\}.
\end{align*}
Similarly, we can provide one-step estimators for $\theta^+_{0,\tau_0}(f)$ and $\theta^{-}_{0,\tau_0}(f)$:
\begin{align*}
&\theta^+_{n,\tau_n}(f) := \frac{1}{n}\sum_{i=1}^n\left\{\psi_n(Z_i) -\frac{1}{n}\sum_{j=1}^n \psi_n(Z_j)\right\}\left\{f(X_{s,i}) - \frac{1}{n} \sum_{j=1}^n f(X_{s,j})\right\}
\end{align*}
and $\theta^-_{n,\tau_n}(f):=-\theta^+_{n,\tau_n}(f)$. We then propose to estimate $\sup_{f \in \mathcal{F}} \theta^+_{0,\delta}(f)$ and $\inf_{f \in \mathcal{F}} \theta^-_{0,\delta}(f)$ as $\sup_{f \in \mathcal{F}} \theta^+_{n,\delta}(f)$ and $\inf_{f \in \mathcal{F}} \theta^-_{n,\delta}(f)$ respectively. Likewise, $\sup_{f \in \mathcal{F}} |\theta^{+}_{0,\tau_0}(f) - \theta^{-}_{0,\tau_0}(f)|$ can be estimated as $\sup_{f \in \mathcal{F}} |\theta^{+}_{n,\tau_n}(f) - \theta^{-}_{n,\tau_n}(f)|=\sup_{f \in \mathcal{F}} |2\theta^{+}_{n,\tau_n}(f) |$.
Whilst calculating the supremum and infimum may appear challenging when $\mathcal{F}$ is an infinite dimensional function class, in Section \ref{sec:imp} we discuss how this can be efficiently done using optimization software. 

\subsection{Constructing the test}\label{sec:construct}

Our proposed test for quantitative effect heterogeneity is of the typical form
\begin{align*}
\phi(Z_1, \ldots, Z_n) :=
    \begin{cases}
        \text{``Do not reject''} & \text{ if } \sup_{f \in \mathcal{F}} |\theta^+_{n,\tau_n}(f) - \theta^-_{n,\tau_n}(f)|  \leq n^{-1/2}t_\alpha 
        \\
        \text{``Reject''} & \text{ if } \sup_{f \in \mathcal{F}} |\theta^+_{n,\tau_n}(f) - \theta^-_{n,\tau_n}(f)|  > n^{-1/2}t_\alpha
    \end{cases},
\end{align*}
where $t_{\alpha}$ is chosen so that the asymptotic type I error rate is controlled at the level $\alpha$. Similarly, to ensure that our proposed test for qualitative heterogeneity asymptotically controls the type I error level, we perform two one-sided tests of the null hypotheses that $\sup_{f \in \mathcal{F}} \theta^+_{0,\delta}(f) \leq 0$ and $\inf_{f \in \mathcal{F}} \theta^-_{0,\delta}(f) \geq 0$. In particular, we consider a test of the form
\begin{align*}
\phi(Z_1, \ldots, Z_n) :=
    \begin{cases}
        \text{``Do not reject''} & \text{ if } \sup_{f \in \mathcal{F}} \theta^+_{n,\delta}(f) \leq n^{-1/2}t^+_\alpha  \textbf{ or } \inf_{f \in \mathcal{F}} \theta^-_{n,\delta}(f) \geq n^{-1/2}t^-_\alpha
        \\
        \text{``Reject''} & \text{ if } \sup_{f \in \mathcal{F}} \theta^+_{n,\delta}(f) > n^{-1/2}t^+_\alpha \textbf{ and } \inf_{f \in \mathcal{F}} \theta^-_{n,\delta}(f) < n^{-1/2}t^-_\alpha
    \end{cases},
\end{align*}
where $t^+_\alpha$ and $t^-_\alpha$ are selected to control the type I error rate at the level $\alpha$. To choose threshold values for each of the above tests that ensure asymptotic size control requires some knowledge of the distributions of the tests statistics under the null hypothesis. In Section \ref{sec:theory}, we will show that $n^{1/2}\sup_{f \in \mathcal{F}}|\theta^+_{n,\tau_n}(f) - \theta^-_{n,\tau_n}(f)-\{\theta^+_{0,\tau_0}(f) - \theta^-_{0,\tau_0}(f)\}|$, $n^{1/2}\sup_{f\in\mathcal{F}}\{\theta_{n,\delta}^+(f)-\theta_{0,\delta}^+(f)\}$, and $n^{1/2}\inf_{f \in \mathcal{F}}\{\theta_{n,\delta}^-(f)-\theta_{0,\delta}^-(f)\}$ all converge weakly to the supremum (or infimum) of a Gaussian process. As a result, when $t_\alpha$ is selected as the $1-\alpha$ quantile of the limiting distribution of $n^{1/2}\sup_{f \in \mathcal{F}}|\theta^+_{n,\tau_n}(f) - \theta^-_{n,\tau_n}(f)-\{\theta^+_{0,\tau_0}(f) - \theta^-_{0,\tau_0}(f)\}|$, the quantitative test achieves nominal type I error control.
For the qualitative tests, the type I error rate is controlled by selecting $t^+_\alpha$ as the $1-\alpha$ quantile of $n^{1/2}\sup_{f\in\mathcal{F}}\{\theta_{n,\delta}^+(f)-\theta_{0,\delta}^+(f)\}$, and selecting $t^-_\alpha$ as the $\alpha$ quantile of $n^{1/2}\inf_{f \in \mathcal{F}}\{\theta_{n,\delta}^-(f)-\theta_{0,\delta}^-(f)\}$. However, a closed-form representation of the relevant asymptotic distributions will not generally be available. 

We will therefore use an approximation of a Gaussian process based on the multiplier bootstrap method, derived from the multiplier central limit theorem \citep{vanderVaart1996}. For $m = 1,\ldots, M$ and $M$ large, let $\xi^m_1, \ldots, \xi^m_n$ be a random sample of independent Rademacher random variables (also independent of $Z$).
Let $T_m$, $T_m^+$ and $T_m^-$ be given by
\begin{align*}
    &T_m := \sup_{f \in \mathcal{F}}\bigg|n^{-1/2}\sum_{i=1}^n \xi_i\bigg[\left\{\psi_n(Z_i) - \frac{1}{n}\sum_{j=1}^n \psi_n(Z_j) \right\} \left\{2f(X_{s,i}) - \frac{2}{n}\sum_{j=1}^n f(X_{s,j}) \right\} \\&\quad \quad \quad  \quad \quad  - \left\{ \theta^+_{n,\tau_n}(f) - \theta^-_{n,\tau_n}(f) \right\} \bigg] \bigg|
    \\
    &T_m^+ := \sup_{f \in \mathcal{F}} n^{-1/2}\sum_{i=1}^n \xi_i^m \left[\left\{\psi_n(Z_i) - \delta \right\} f(X_{s,i}) - \theta^+_{n,\delta}(f)\right],
    \\
    &T_m^- := \inf_{f \in \mathcal{F}} n^{-1/2}\sum_{i=1}^n \xi_i^m\left[\left\{\psi_n(Z_i) - \delta \right\} \left\{1 - f(X_{s,i})\right\} - \theta^-_{n,\delta}(f) \right].
\end{align*}
 By making $M$ large, the conditional distribution of $T_{1},\ldots,T_{M}$ given $Z_1, \ldots, Z_n$ approximates the limiting distribution $T$ of $n^{1/2}\sup_{f \in \mathcal{F}}|\theta^+_{n,\tau_n}(f) - \theta^-_{n,\tau_n}(f)-\{\theta^+_{0,\tau_0}(f) - \theta^-_{0,\tau_0}(f)\}|$. Similarly, the conditional distributions of $T_{1}^+,\ldots,T_M^+$ and $T_{1}^-,\ldots T_M^-$, given the data, approximate the limiting distributions $T^+$ and $T^-$ of $n^{1/2}\sup_{f\in\mathcal{F}}\{\theta_{n,\delta}^+(f)-\theta_{0,\delta}^+(f)\}$ and $n^{1/2}\inf_{f \in \mathcal{F}}\{\theta_{n,\delta}^-(f)-\theta_{0,\delta}^-(f)\}$ respectively. Specifically, for large enough $M$, one can approximate $t_\alpha$, $t^+_\alpha$, and $t^-_\alpha$ as the $(1-\alpha)$ quantile of $T_1,\ldots,T_M$, the $(1-\alpha)$ quantile of $(T^+_1, \ldots, T^+_M)$ and the $\alpha$ quantile of $(T^-_1, \ldots, T^-_M)$, respectively. A summary of our methodology is given in Algorithms \ref{algo-quant} and \ref{algo-qual}. In our simulations and data analysis, we chose $M=2000$ and $M = 10,000$, respectively.

\begin{algorithm}
Estimate nuisance parameters $\mu_0(a,x)$ and $\pi_0(a|x)$ as $\mu_n(a,x)$ and $\pi_n(a|x)$.\\
Select $\mathcal{F}$; estimate $\theta^+_{0,\tau_0} - \theta^-_{0,\tau_0}$ as $\sup_{f\in\mathcal{F}}|\theta^+_{n,\tau_n}(f) - \theta^-_{n,\tau_n}(f)|$.\\
Use the multiplier bootstrap to generate the empirical distribution of $n^{1/2}\sup_{f \in \mathcal{F}}|\theta^+_{n,\tau_n}(f) - \theta^-_{n,\tau_n}(f)-\{\theta^+_{0,\tau_0}(f) - \theta^-_{0,\tau_0}(f)\}|$.\\
Select $t_{n,\alpha}$ as the $(1-\alpha)$ quantile of $(T_1, \ldots, T_M)$.\\
Reject the null hypothesis if $n^{1/2}\sup_{f\in\mathcal{F}}|\theta^+_{n,\tau_n}(f) - \theta^-_{n,\tau_n}(f)| > t_{n,\alpha}$.
\caption{Construction of hypothesis test for quantitative heterogeneity}
\label{algo-quant}
\end{algorithm}

\begin{algorithm}
Estimate nuisance parameters $\mu_0(a,x)$ and $\pi_0(a|x)$ as $\mu_n(a,x)$ and $\pi_n(a|x)$. 
\\
Select $\mathcal{F}$; estimate $\theta^+_{0,\delta}$ and $\theta^-_{0,\delta}$ as $\sup_{f\in\mathcal{F}}\theta^+_{n,\delta}(f)$ and $\inf_{f\in\mathcal{F}}\theta^-_{n,\delta}(f)$ respectively.\\
Use the multiplier bootstrap to generate the empirical distributions of $n^{1/2}\sup_{f\in\mathcal{F}}\{\theta_{n,\delta}^+(f)-\theta_{0,\delta}^+(f)\}$ and $n^{1/2}\inf_{f \in \mathcal{F}}\{\theta_{n,\delta}^-(f)-\theta_{0,\delta}^-(f)\}$.\\
Select $t^+_\alpha$ and $t^-_\alpha$ as the $(1-\alpha)$ quantile of $(T^+_1, \ldots, T^+_M)$ and the $\alpha$ quantile of $(T^-_1, \ldots, T^-_M)$, respectively.\\
Reject the null hypothesis if $n^{1/2}\sup_{f \in \mathcal{F}} \theta^+_{n,\delta}(f) > t^+_{n,\alpha} \textbf{ and } n^{1/2}\inf_{f \in \mathcal{F}} \theta^-_{n,\delta}(f) < t^-_{n,\alpha}$.
\caption{Construction of hypothesis test for qualitative heterogeneity}
\label{algo-qual}
\end{algorithm}

Whilst our paper has focused on estimation and testing, a related problem is constructing confidence intervals for our estimands. Because these parameters are constrained to be either non-negative or non-positive, interval construction via a Gaussian approximation is ill-advised. In theory, valid intervals can be constructed via the inversion of tests.

\section{Theoretical details}\label{sec:theory}

\subsection{Asymptotic distribution of estimators}

The efficient influence function is key to obtaining an asymptotically linear representation of our estimators. In order to show this, we first require some additional assumptions.  
\begin{assumption}{(Donsker class)}\label{donsker}
$\varphi^+_{0,\delta}(\cdot;f)$, $\varphi^-_{0,\delta}(\cdot;f)$, $\varphi^+_{0,\tau_0}(\cdot;f)$, and $\varphi^-_{0,\tau_0}(\cdot;f)$  all lie in a $P_0$-Donsker class for each $f \in \mathcal{F}$. The same holds for the estimated $\varphi^+_{n,\delta}(\cdot;f)$, $\varphi^-_{n,\delta}(\cdot;f)$, $\varphi^+_{n,\tau_n}(\cdot;f)$ and $\varphi^-_{n,\tau_n}(\cdot;f)$ with probability tending to one.
\end{assumption}
\begin{assumption}{(Consistency of nuisance parameter estimators)}\label{consist} For each $a\in \{0,1\}$,
\begin{align*}
\int \{\mu_n(a,x)-\mu_0(a,x)\}^2dP_0(x)=o_{P_0}(1),\\
\int \{\pi_n(a|x)-\pi_0(a|x)\}^2dP_0(x)=o_{P_0}(1).
\end{align*}
\end{assumption}
\begin{assumption}{(Product rate condition)}\label{product} For each $a\in \{0,1\}$,
\begin{align*}
\left[\int \{\mu_n(a,x)-\mu_0(a,x)\}^2dP_0(x)\right]^{1/2}
\left[\int \{\pi_n(a|x)-\pi_0(a|x)\}^2dP_0(x)\right]^{1/2}=o_{P_0}(n^{-1/2}).
\end{align*}
\end{assumption}
Assumption \ref{donsker} can be justified if both $i)$ $\mathcal{F}$ is a $P_0$-Donsker class and $ii)$ the estimated nuisance parameters also fall within a $P_0$-Donsker class. This follows from preservation property of Donsker classes; see Lemma S1 in \citet{hudson2026inference}. We emphasize that $i)$ is not a condition on the true CATE, but rather on the complexity on the class of rules.  If the optimal rule does not fall within the class $\mathcal{F}$, this will not jeopardize type I error. We do no expect $i)$ to be too limiting in practice, since making $\mathcal{F}$ very `large' could lead to high variability in the test statistic and lead to infeasible optimization procedures. See  Section \ref{sec:imp} for examples of $\mathcal{F}$ that satisfy this condition. 

Condition $ii)$ in the previous paragraph would restrict the complexity of $\mu_n(a,x)$ and $\pi_n(a|x)$. Many parametric and non-parametric estimators fulfill this condition. We conjecture that condition $ii)$ could be weakened using cross-fitting \citep{chernozhukov2018double}. Moreover, when the propensity score is known, one can relax the Donsker conditions and allow for nuisance misspecification. Assumption \ref{consist} requires consistency of the nuisance parameter estimators, whilst Assumption \ref{product} allows for one to converge slower, so long as the other converges quickly as $n$ grows. These assumptions are now standard in the literature on de-biased learning of treatment effects \citep{kennedy2024semiparametric}. 

\begin{theorem}\label{theorem:al}
(Asymptotic linearity) Suppose that $\pi_n(a|X)\geq \epsilon$ and $\pi_0(a|X)\geq \epsilon \quad \forall a\in \{0,1\}$ for some $\epsilon>0$ , and $|Y-\mu_n(A,X)|\leq C$ for $C<\infty$, all with probability one. Then if Assumptions \ref{positivity}, \ref{donsker}, \ref{consist} and \ref{product} also hold, $\theta^+_{n,\delta}(f)$ and $\theta^-_{n,\delta}(f)$ admit the following representation:
\begin{align*}
    \theta^+_{n,\delta}(f) - \theta^+_{0,\delta}(f) &= \frac{1}{n}\sum_{i=1}^n \varphi^+_{0,\delta}(Z_i;f) + r^+_{n,\delta}(f),
    \\
    \theta^-_{n,\delta}(f) - \theta^-_{0,\delta}(f) &= \frac{1}{n}\sum_{i=1}^n \varphi^-_{0,\delta}(Z_i;f) + r^-_{n,\delta}(f),\\
    \theta^+_{n,\tau_n}(f) - \theta^+_{0,\tau_0}(f) &= \frac{1}{n}\sum_{i=1}^n \varphi^+_{0,\tau_0}(Z_i;f) + r^+_{n,\tau_n}(f),
    \\
    \theta^-_{n,\tau_n}(f) - \theta^-_{0,\tau_0}(f)&= \frac{1}{n}\sum_{i=1}^n \varphi^-_{0,\tau_0}(Z_i;f) + r^-_{n,\tau_n}(f),
\end{align*}
where $\sup_{f\in \mathcal{F}}|r^+_{n,\delta}(f)|=o_{P_0}(n^{-1/2})$, $\sup_{f\in \mathcal{F}}|r^-_{n,\delta}(f)|=o_{P_0}(n^{-1/2})$, $\sup_{f\in \mathcal{F}}|r^+_{n,\tau_n}(f)|=o_{P_0}(n^{-1/2})$ and $\sup_{f\in \mathcal{F}}|r^-_{n,\tau_n}(f)|=o_{P_0}(n^{-1/2})$.
\end{theorem}
This result establishes in particular the \textit{uniform} asymptotic linearity of our estimators with respect to $\mathcal{F}$.
Pointwise asymptotic linearity for any fixed $f\in \mathcal{F}$ is not sufficient for our purposes, since our tests depend on a supremum/infimum statistic taken over a function class. Uniform consistency of our estimator follows as an immediate consequence of the uniform asymptotic linearity result. The following result states that if, in addition, the function class $\mathcal{F}$ is not overly complex, our estimator also converges weakly to a Gaussian process. This can be seen to hold through an application of Slutsky's theorem; see, e.g., Theorem 7.15 of \citet{kosorok2008introduction}.
\begin{corollary}\label{corollary:weak}
(Weak convergence) Under the conditions of Theorem \ref{theorem:al}, $\{n^{1/2}[\theta^+_{n,\delta}(f) - \theta^+_{0,\delta}(f)]:f\in \mathcal{F}\}$ converges weakly to a tight Gaussian process $\mathbb{G}^+$ as an element of $l^\infty(\mathcal{F})$,  where $l^\infty(\mathcal{F})$ is the vector space of bounded real-valued functionals on $\mathcal{F}$. Here, $\mathbb{G}^+$ has mean zero and covariance $\Sigma^+:(f_1,f_2)\mapsto E_0[\varphi^+_{0,\delta}(Z;f_1)\varphi^+_{0,\delta}(Z;f_2)]$. Similarly $\{n^{1/2}[\theta^-_{n,\delta}(f) - \theta^-_{0,\delta}(f)]:f\in \mathcal{F}\}$ converges weakly to a tight Gaussian process $\mathbb{G}^-$, where $\mathbb{G}^-$ is equivalently defined.
Finally, $\{n^{1/2}[\{\theta^+_{n,\tau_n}(f) - \theta^-_{n,\tau_n}(f)\} - \{\theta^+_{0,\tau_0}(f) - \theta^-_{0,\tau_0}(f)\}]: f \in \mathcal{F}\}$ converges weakly to a tight Gaussian process $\mathbb{G}$ with mean zero and covariance $\Sigma: (f_1, f_2) \mapsto E_0[\{\varphi^+_{0,\tau_0}(Z;f_1)-\varphi^-_{0,\tau_0}(Z;f_1)\}\{\varphi^+_{0,\tau_0}(Z;f_2) - \varphi^-_{0,\tau_0}(Z;f_2)\}]$.
\end{corollary}

In the supplementary material, we convert Theorem \ref{theorem:al} and Corollary \ref{corollary:weak} into formal results on type I error control and power. For fixed null and alternative hypotheses, in Appendix \ref{appendix:fixed} we establish type I error control and consistency for both the quantitative and qualitative tests. In Appendix \ref{appendix:local}, we also study the properties of our tests in a \textit{local asymptotic} setting, wherein the data-generating distribution changes with $n$. The qualitative tests have non-trivial power when there is a strong beneficial effect in certain subgroups and a weak harmful effect in others (or vice versa). However, we cannot make guarantees when the beneficial and harmful effects are both weak. 

Furthermore, $\theta^+_{n,\delta}(f)$ ($\theta^-_{n,\delta}(f)$) can be used individually to test the null that the CATE is less than or equal (greater than or equal) to $\delta$. This could be of practical interest, for example, when practitioners aim to detect if certain subgroups have a practically (clinically) significant effect. These tests would have non-trivial power against local alternatives, comparable to the procedures of \citet{hsu2017consistent} and \citet{shi2019sparse}.

\section{Implementation}\label{sec:imp}

\subsection{Choice of $\mathcal{F}$}

\noindent \textbf{Approach 1: Linear Threshold Rules.} Consider the class of indicator functions of whether a linear function of $x_s$ is non-negative:
\begin{align*}
    \mathcal{F}:=& \left\{ f:x_s \mapsto \mathds{1}(\rho_0+\rho_1^Tx_s \geq  0): (\rho_0,\rho_1) \in \mathbb{R}^{|s|+1} \right\}. 
\end{align*}
This class is familiar from the literature on optimal treatment regimes \citep{kitagawa2018should}, where it is popular due to the transparency and interpretability of linear rules. It follows from Theorems 4.2.1 and 10.1.4 of \citet{dudley2014uniform} that this class has finite VC dimension and because it consists of indicator functions, it hence satisfies our Assumption \ref{donsker}. 
In the case that $X_s$ is scalar, one may consider the further simplification 
\begin{align}\label{class:consistant_indicator}
    \mathcal{F}:= \left\{ f:x_s \mapsto \mathds{1}(x_s \geq x_{0,s}): x_{0,s} \in \mathbb{R} \right\} \cup \left\{ f:x_s \mapsto \mathds{1}(x_s \leq x_{0,s}): x_{0,s} \in \mathbb{R} \right\}
\end{align}
where $\mathcal{F}$ includes indicators of whether $X_s$ exceeds a given threshold. It can be seen that whenever the CATE curve $\tau_{0,s}$ is monotone, $\mathcal{F}$ is correctly-specified in the sense that it contains $g_{0,\delta}$ in \eqref{opt-rule} for any $\delta$.
Thus, we have equality in \eqref{quant-ineq} and \eqref{qual-ineq}.

\noindent \textbf{Approach 2: Bounded Variation.} 
Let $|s| = 1$; for a large positive integer $p$, let $\tilde{x}_{s,1} \leq \ldots \tilde{x}_{s,p}$ define a grid on $\mathbb{R}$.
For $k \in \{1, \ldots, p\}$, we define the $(k+1)$-th basis function as $h_{k+1}: x_s \mapsto \mathds{1}(\tilde{x}_{s,k-1} \leq x_s \leq \tilde{x}_{s,k})$, and we also define $h_1: x_s \mapsto \mathds{1}(x_s \leq \tilde{x}_{s,1})$. We then set $\mathcal{F}$ as
\begin{align*}
    \mathcal{F}:= \left\{f: x_s \mapsto \left(\sum_{k=1}^p b_k h_k(x_s)\right): b_1, b_2, \ldots \in \mathbb{R}, 
    \sum_{k = 1}^{p-1} |b_{k+1} - b_k| \leq \lambda, f\in[0,1] \right\}
\end{align*}
for some $0<\lambda < \infty$. 
The class $\mathcal{F}$ contains functions with  total variation norm bounded above by $\lambda$, which modulates the complexity of the class. For sufficiently large $\lambda$ and $p$, $\mathcal{F}$ will contain a close approximation of $g_{0,\delta}$. Note that $\lambda=1$ if $\mathcal{F}$ captures all monotone functions and $\lambda=2$ if it also captures all convex/concave functions. In Appendix \ref{appendix:bonus}, we describe a more general approach based on basis expansions.

We note that although the optimal choice of $f$ over all classes will be a 0-1 rule, Approach 2 leaves the possibility of returning $f$ in $(0,1)$ (corresponding to a stochastic rule). This is not contradictory because $\mathcal{F}$ need not include the optimal rule. 
In fact, we believe using a function class that contains stochastic rules can have advantages in some settings e.g. when the CATE curve has many roots. 
Moreover if the conditional treatment effect is small relative to the sample size for many subgroups, estimation of the optimal rule will be difficult. By considering stochastic rules, we allow for (nearly) deterministic  decisions to be made for subgroups where the CATE is large, whereas stochastic decisions can be made for subgroups for which the CATE is relatively close to zero, and it is difficult to determine in a small sample size whether the treatment is beneficial or detrimental.

\noindent{\textbf{Approach 3: Finite-Depth Trees}}
A depth-0 decision tree $D_0$ is a rule $D_0(x_s)=\alpha$ where $\alpha\in \{0,1\}$. For any $1\leq K< \infty$, a depth-$K$ tree $D_K$ is constructed using a splitting variable $j\in 1,...,|s|$, a threshold $t\in \mathbb{R}$ and two depth-$(K-1)$ decision trees $D_{(K-1),i}$ and $D_{(K-1),ii}$ \citep{athey2021policy}. Specifically, $D_K(x_s)=D_{(K-1),i}$ if $x_{j}\leq t$ and $D_K(x_s)=D_{(K-1),ii}$ otherwise. When the covariate dimension and depth of the decision trees are fixed, then the VC dimension of the class is finite. 

In practice, one may want to use the data to select $\mathcal{F}$. It follows from Theorem 2 in \citet{hudson2026inference} that this can be done using cross-validation without compromising type I error, so long as the estimated choice of $\mathcal{F}$ converges to a fixed class. However, empirical results suggest that type I error inflation can occur with data-driven tuning parameter selection. This could be reduced by sample-splitting, at the cost of a potential penalty in power. To perform inference on the maximum of a high-dimensional vector of Studentized statistics, \citet{chernozhukov2019inference} propose a two-step screening method to first reduce the length of the vector, which does not require splitting. It relies on theory for non-Donsker random processes, and it would be an interesting direction to extend this approach to our setting where $\mathcal{F}$ may be infinite-dimensional.

As we will see in the simulations, structural knowledge of $\mathcal{F}$ improves performance. Hence $\mathcal{F}$ should be chosen to reflect any \textit{a priori} knowledge on the CATE. For example, if it is monotone increasing in a scalar $X_s$, then one might choose simple linear threshold rules (Approach 1) or bounded variation with $\lambda=1$. Otherwise, we recommend choosing more flexible classes e.g. more flexible threshold rules (see the supplementary material Appendix \ref{appendix:simulations}) or decision trees, which have reasonable power against a range of alternatives.

\subsection{Computation}

Approaches 1 and 3 can be implemented using mixed integer programming \citep{kitagawa2018should,athey2021policy}. For the threshold class \eqref{class:consistant_indicator}, one can recalculate $\theta^+_{n,\delta}(f)$ and $\theta^-_{\delta,f,n}$ over all cut-offs defined by the observed values of $X_s$, and take the maximum/minimum. For large $n$, one could instead work with a reduced grid of values. For Approach 2,  the optimisation problem can be solved using standard software for convex programming e.g. \verb|CVXR| in \verb|R| \citep{fu2020cvxr}. The dimension of the basis $p$ can in principle be large, although this must be traded off with the computational complexity. This choice of $\mathcal{F}$ grants a greater degree of flexibility as the class contains functions that are discontinuous and can be locally non-smooth. In order to respect the bounds on $f$, we also recommend re-scaling each basis function so that it falls in $[0,1]$ and enforcing the constraint that the coefficients also reside within [0,1]. Developing feasible computational routines for other function classes is an important topic for future work.

\section{Simulations}\label{sec:sims}

\subsection{Simulation design}\label{sec:sims_design}

Let $X = (X_1, X_2, X_3)$ be a vector of independent uniform random variables on the interval $[-1, 1]$.
We generate the treatment assignment variable $A$ from the conditional distribution $\pi_0(1|x) = \text{expit}\left\{\frac{1}{8}x_1 + \frac{1}{4}\sin(\pi x_2) \right\}$. Given $X$ and $A$, we generate the outcome as $Y = h(X_1, X_2, X_3) + A \gamma(X_3) + \epsilon$,
where $h(x) = x_1 + \text{expit}\left\{\frac{1}{2}\left(x_2 + x_3\right)\right\}$, $\epsilon \sim N(0, 3^2)$ is white noise, and  $\gamma$ is a function of $X_3$ that we manipulate.
It can be seen that the CATE, given $X_3$, is equal to $\gamma$. Our simulations assess various tests' performance for assessing heterogeneity in $X_3$; see supplementary material Appendix \ref{appendix:simulations} for multivariable heterogeity.

We consider the following specifications of $\gamma$ to control whether the quantitative and/or qualitative nulls hold:
\begin{itemize}
    \item Setting 1 (No heterogeneity)
    $\gamma(x_3) = \frac{3}{4}$.
    \item Setting 2 (Quantitative heterogeneity; monotone CATE): $\gamma(x_3) = 15(x_3 - 0.5)\mathds{1}(x_3 > 0.5)$.
    \item Setting 3 (Quantitative heterogeneity; non-monotone CATE): $\gamma(x_3) = 3(1 - x_3^2)$.
    \item Setting 4 (Qualitative heterogeneity; monotone CATE): $\gamma(x_3) = 3\text{sign}(x_3) x_3^2$.
    \item Setting 5 (Qualitative heterogeneity; non-monotone CATE): $\gamma(x_3) = 3\cos \left(\frac{3\pi}{2}x_3\right)$.
\end{itemize}

We consider the following methods (abbreviated names from Figures \ref{fig:sims-quant-null}-\ref{fig:sims-qual-alt} are in quotes):
\begin{itemize}
\item Quant (Qual) Monotone: our proposed quantitative (qualitative) test implemented using Approach 1 in Section \ref{sec:imp} (`Mono').
\item Quant (Qual) Non-monotone: our quantitative (qualitative) test using Approach 2, with $\lambda=2$ (`Non-mono').
\item Quant (Qual) Kernel ridge: Kernel ridge regression is used to obtain a doubly robust (DR)-learning estimator for the CATE \citep{kennedy2023towards}, which is used to form a plug-in estimator for $\theta^+_{0, \tau_0}$ and $\theta^-_{0, \tau_0}$ ($\theta^+_{0, \delta}$ and $\theta^-_{0, \delta}$); see Appendix \ref{appendix:simulations} for details (`Ridge').
\item Quant (Qual) Sample-split: The independent observations are randomly split into two partitions with equal probability. On the first partition, kernel ridge regression is used to obtain a DR-learning estimator for the CATE. The CATE estimator is then used to form a plug-in estimator for the optimal decision rule $g_{0,\tau_0}$ ($g_{0, \delta}$). On the second partition, our proposed test is performed using the singleton set $\mathcal{F} = \{g_{0, \tau_0}\}$ ($\mathcal{F} = \{g_{0, \delta}\}$). See Appendix \ref{appendix:simulations} for details (`Split').
\item Quant Unstructured: discretize $X_3$ into $k$ equally-spaced intervals. Estimate the ATE and each subgroup effect using augmented inverse probability weighting. The test statistic is the sum of the absolute differences between the subgroup estimates and the ATE (`Unstr').
\item Qual Gail-Simon: the test of \cite{gail1985testing} with $k$ subgroups (`GS').
\item Qual Range: the range test of \cite{piantadosi1993comparison} with $k$ subgroups  (`Range').
\end{itemize}
The conditional mean and propensity score are estimated using the highly adaptive lasso, a flexible nonparametric regression method \citep{benkeser2016highly}. To implement our approaches, we divide the interval $[-1,1]$ into $20$ or $50$ equally-spaced sub-intervals. We therefore chose $k\in\{20, 50\}$ for the comparator tests, which reflect the standard strategy of discretizing covariates and testing for differences between the constructed subgroups. For Quant Unstructured, since the difference between subgroup estimators and the ATE estimator is jointly normal with zero mean under the null, the null limiting distribution of the test statistic can be approximated using Monte Carlo sampling. All tests were performed at the nominal level $\alpha = .05$. Under each of the above settings, we generated 1000 random data sets for $n \in \{250, 500, 1000, 2000\}$. 

\subsection{Simulation results}

\begin{figure}
    \centering
    \includegraphics[width=0.75\linewidth]{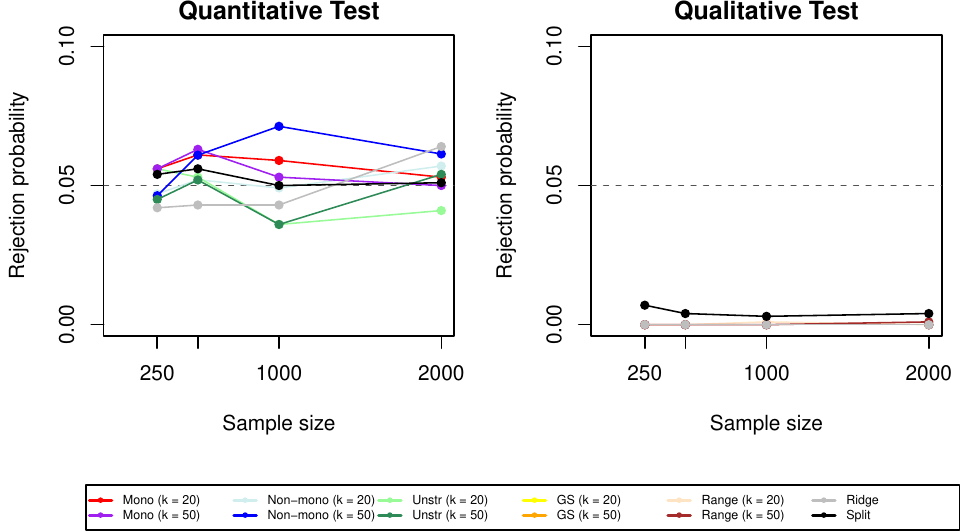}
    \caption{Monte Carlo estimate of rejection probability, when the null hypothesis of no quantitative or qualitative qualitative heterogeneity holds.  The definitions of the abbreviations in the legend are given  in Section \ref{sec:sims_design}}. 
\label{fig:sims-quant-null}
\end{figure}

\begin{figure}
    \centering
    \includegraphics[width=0.75\linewidth]{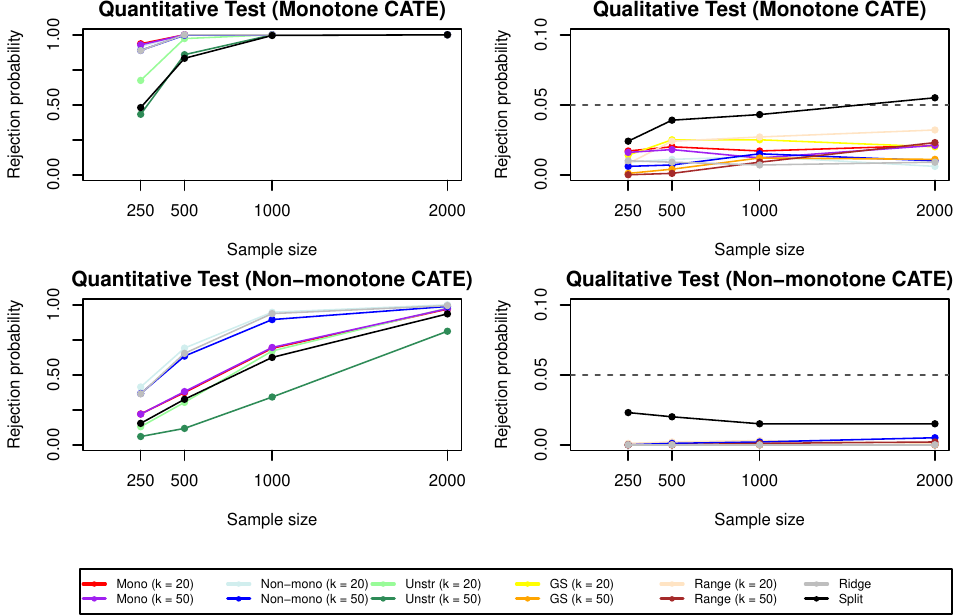}
    \caption{Monte Carlo estimate of rejection probability in the presence of quantitative heterogeneity and the absence of qualitative heterogeneity.} 
\label{fig:sims-quant-alt}
\end{figure}

\begin{figure}
    \centering
    \includegraphics[width=0.75\linewidth]{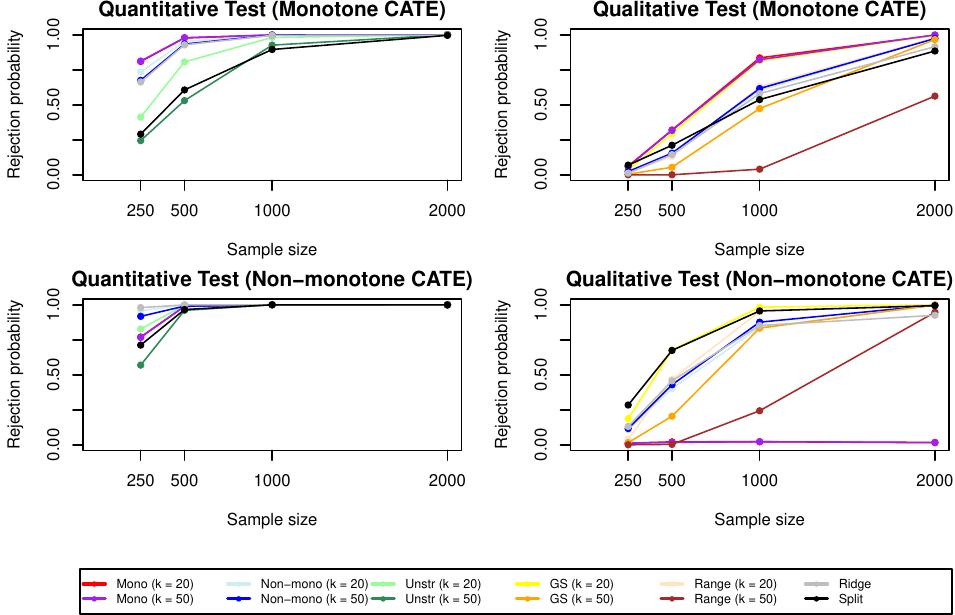}
    \caption{Monte Carlo estimate of rejection probability in the presence of quantitative heterogeneity and qualitative heterogeneity.} 
\label{fig:sims-qual-alt}
\end{figure}

Figure \ref{fig:sims-quant-null} shows Monte Carlo estimates of rejection probabilities when the null holds (Setting 1).
For the quantitative test, tight type I error control is achieved by all methods for $n$ large enough. Figure \ref{fig:sims-quant-alt} shows rejection probabilities in the setting where quantitative interactions occur but qualitative interactions do not occur. For all quantitative tests, power approaches one as the sample size increases, as expected. Approach 1 (Quant Monotone) performs best when the CATE is monotone, and the performance is nearly identical whether $k = 20$ or $k = 50$ thresholds are used. When the CATE is quadratic, Approach 2 with a either a small or moderate number of basis functions (Quant Non-monotone; $k = 20$ or $k = 50$) and the kernel ridge regression method  perform best. The comparator (Quant Unstructured), which does not make use of structural assumptions on the CATE, is less powerful than both implementations of our proposal when a large number of bins ($k = 50$) is used, though it performs comparably to the monotone approach when a conservative number of bins is used $(k = 20)$. It relies on estimated treatment effects from the $k$ constructed subgroups, which have high variances when the total sample size is small. The sample-split test was also generally less powerful than our proposal. For multivariable heterogeneity, larger power gains were observed for our proposal versus the sample-split and kernel ridge tests (Appendix \ref{appendix:simulations}). All qualitative tests have type I error between 0 and $\alpha$ for all $n$. In certain cases, we see that the rejection probability increases with $n$. This occurs because the one-sided test for a positive effect is likely rejecting with probability tending to one, whereas the test for a negative effect has type I error rate between $\alpha$ and zero.

Figure \ref{fig:sims-qual-alt} displays Monte Carlo estimates of rejection probabilities in the presence of qualitative interactions.
The quantitative tests perform similarly as in the previous setting.
Among the qualitative tests, our approach with implementation Approach 1 (Qual Monotone) performs best when the CATE is monotone, and similar performance is achieved irrespective of the number of thresholds $k$.  The Gail-Simon test achieves comparable performance with $k = 20$ bins, but power declines as the number of subgroups increases to $k = 50$. When the CATE is non-monotone, the Gail-Simon test with $k = 20$ bins and the sample-split test perform best. Approach 2 (Qual Non-monotone) has competitive performance in this setting, though is less powerful than the aforementioned two approaches. While the power of the Gail-Simon test decreases as $k$ increases, Approach 2 has similar power in either case. The range test (Qual Range) has moderate power with $k = 20$, and power greatly declines when $k = 50$.
Of note, power against the qualitative alternative is generally much lower than power against the quantitative alternative.
This occurs because the qualitative null is composite and hence generally more difficult to reject.

\section{Data analysis}\label{sec:data}

We demonstrate our proposed methodology by analyzing data from the AIDS Clinical Trail Group (ACTG) Study 175 \citep{hammer1996trial}\footnote{The data are available at \url{https://archive.ics.uci.edu/dataset/890/aids+clinical+trials+group+study+175.}}. 
This was a randomized trial which compared treatments for human immunodeficiency virus type I (HIV) in an adult population with CD4 count between 200 to 500 per cubic millimeter.
The following treatments were considered: goal of comparing monotherapy with zidovudine ($A = 0$) versus monotherapy with didanosine, combination therapy with zidovudine and didnosine, or combination therapy with zidovudine and zalcitabine ($A = 1$).
Our analysis studies the effect of the treatment on the composite outcome of occurrence of a fifty percent decline in the CD4 cell count, development of the acquired immunodeficiency syndrome (AIDS), or death. We treat the outcome as a binary indicator of the event occurring within two years. After omitting from the sample study participants for whom events were censored before two years, we retained a sample of $n = 1,938$.

\begin{figure}
    \centering
    \includegraphics[width=0.75\linewidth]{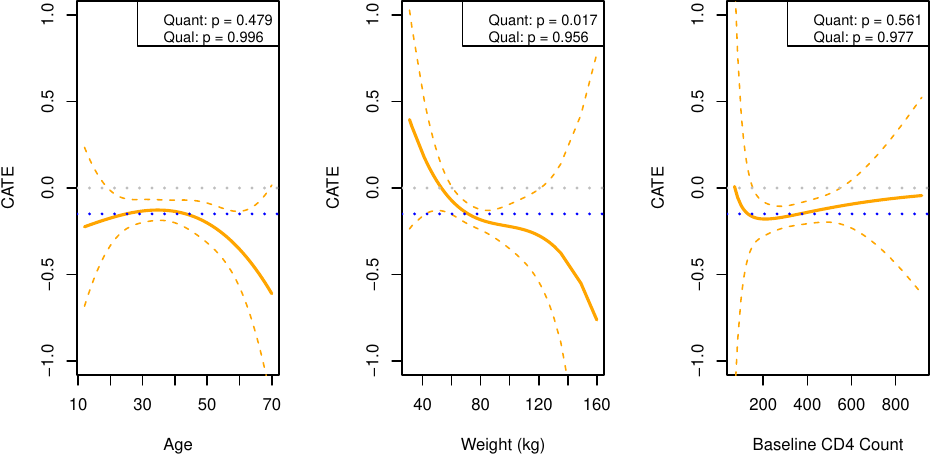}
    \caption{Cubic spline estimates of CATE curves and p-values from tests of treatment effect heterogeneity, for the ACTG data. Dashed orange lines represent pointwise 95\% confidence intervals. Dashed grey and blue lines pass through zero and the ATE respectively. Reported p-values for the qualitative tests are taken as the maximum of the individual p-values for one-sided tests for positive and negative effects.}
    \label{fig:actg}
\end{figure}

We assess quantitative and qualitative heterogeneity using weight, age, and baseline $\log_{10}$ CD4 count.
For visualization purposes, crude parametric estimates and pointwise 95\% confidence intervals for the CATE curves, given each covariate, are obtained by regressing $Y(2A - 1)$ on a given covariate in a finite-dimensional cubic spline model.
Quantitative and qualitative tests for heterogeneity are preformed using Approach 1; see Figure \ref{fig:actg} for results.
There is no evidence of quantitative heterogeneity depending on age (p = 0.479) or baseline CD4 count (p = 0.561), though there is modest evidence for quantitative heterogeneity by weight (p = 0.017).
We also performed a simultaneous test for quantitative heterogeneity using the linear threshold rule approach described in the Supplementary Materials.
This test yielded no significant evidence for heterogeneity (p = 0.285).
We find no evidence of qualitative heterogeneity using any covariate, though this is unsurprising as qualitative heterogeneity is difficult to detect in the absence of very strong signal.
We observed similar results when performing a simultaneous test using linear threshold rules (p = 0.999).

Based on these results, the weak evidence of quantitative heterogeneity due to weight suggests that a dynamic rule using weight might have a different population impact to adopting a non-dynamic rule. This is potentially informative given that clinicians may choose treatment based on weight. The limited general evidence of heterogeneity is in itself arguably useful, as it emphasizes the uncertainty surrounding the impact of tailored treatment regimes in this context.

\section{Discussion}\label{sec:discuss}

We have proposed a general nonparametric framework for testing effect heterogeneity. As is made rigorous in Appendix \ref{appendix:local}, our tests have a local $n^{-1/2}$-rate of convergence in certain directions. However, it may be possible to construct tests with slower rates of convergence, but which are spread over a wider class of alternatives. If there are small subgroups where the CATE sharply deviates from some $\delta$, then our test may be dominated in terms of power. Nevertheless, this might be a setting where heterogeneity exists but targeting the intervention would not yield a substantive impact at the population level. Our power guarantees (in certain directions) are important given that randomized trials are not often designed to detect heterogeneity. 
If the population on which the test is performed differs substantially from the one to whom the intervention would be given, then it may be advantageous to marginalise over a different distribution of $X$. One can also target covariate distributions for which treatment effects can be learnt more precisely. 
We conjecture that this could lead to more powerful tests in observational data; an alternative strategy based on variance weighting is described in Appendix \ref{appendix:var_weight}.

A limitation of our test for qualitative heterogeneity is that it may suffer low power for alternatives where many treatment effects are weak but non-null. This is a trait shared by many traditional tests for qualitative heterogeneity. Although our test performed well relative to these approaches in simulations, they are still recommended for large trials and observational studies. In the case of testing for qualitative heterogeneity in a trial with a single dichotomous covariate, \citet{zelterman1990tests} shows how one can enlarge the rejection region of the likelihood ratio test in a way that increases power whilst preserving type I error control. The resulting tests have unusual and counter-intuitive properties.

Finally, we note that other estimands could be utilized for testing quantitative/qualitative homogeneity. Many of these are expected to share the boundary null issue, and that development of valid hypothesis tests would hence be an subject for separate work. All summary measures come with benefits and drawbacks; however, ours were specifically chosen to correspond to policy relevance, which we consider to be an advantage in terms of interpretation and implementation. One could alternatively develop a test of quantitative heterogeneity based on the variance of the CATE \citep{sanchez2023robust}; our inferential framework described here could in principle be adapted to this estimand. The parameters considered in our paper have a convenient interpretation in terms of the area above/below the treatment effect curve, and are natural when considering qualitative heterogeneity. On the other hand, working with the variance may more easily lead to a feasible optimization problem for certain function classes \citep{hudson2023nonparametric}. Nonetheless, the resulting tuning parameter choice may be less intuitive than in our framework, and we can leverage optimization techniques from the literature on optimal treatment regimes. It is unclear to us how the power properties would compare in general.

\section{Acknowledgements}

O.D. was supported by the FWO grant 1222522N. M.J.S was supported by the Swiss National Science Foundation.

\bibliographystyle{apalike}
\bibliography{references}

\begin{thebibliography}{}

\bibitem[Athey and Wager, 2021]{athey2021policy}
Athey, S. and Wager, S. (2021).
\newblock Policy learning with observational data.
\newblock {\em Econometrica}, 89(1):133--161.

\bibitem[Benkeser and Van Der~Laan, 2016]{benkeser2016highly}
Benkeser, D. and Van Der~Laan, M. (2016).
\newblock The highly adaptive lasso estimator.
\newblock In {\em 2016 IEEE international conference on data science and advanced analytics (DSAA)}, pages 689--696. IEEE.

\bibitem[Berger, 1989]{berger1989uniformly}
Berger, R.~L. (1989).
\newblock Uniformly more powerful tests for hypotheses concerning linear inequalities and normal means.
\newblock {\em Journal of the American Statistical Association}, 84(405):192--199.

\bibitem[Chang et~al., 2015]{chang2015nonparametric}
Chang, M., Lee, S., and Whang, Y.-J. (2015).
\newblock Nonparametric tests of conditional treatment effects with an application to single-sex schooling on academic achievements.
\newblock {\em The Econometrics Journal}, 18(3):307--346.

\bibitem[Chernozhukov et~al., 2018]{chernozhukov2018double}
Chernozhukov, V., Chetverikov, D., Demirer, M., Duflo, E., Hansen, C., Newey, W., and Robins, J. (2018).
\newblock {Double/debiased machine learning for treatment and structural parameters}.
\newblock {\em The Econometrics Journal}, 21(1):C1--C68.

\bibitem[Chernozhukov et~al., 2019]{chernozhukov2019inference}
Chernozhukov, V., Chetverikov, D., and Kato, K. (2019).
\newblock Inference on causal and structural parameters using many moment inequalities.
\newblock {\em The Review of Economic Studies}, 86(5):1867--1900.

\bibitem[Chernozhukov et~al., 2025]{chernozhukov2025fisher}
Chernozhukov, V., Demirer, M., Duflo, E., and Fern{\'a}ndez-Val, I. (2025).
\newblock Fisher–schultz lecture: Generic machine learning inference on heterogeneous treatment effects in randomized experiments, with an application to immunization in india.
\newblock {\em Econometrica}, 93(4):1121--1164.

\bibitem[Crump et~al., 2008]{crump2008nonparametric}
Crump, R.~K., Hotz, V.~J., Imbens, G.~W., and Mitnik, O.~A. (2008).
\newblock Nonparametric tests for treatment effect heterogeneity.
\newblock {\em The Review of Economics and Statistics}, 90(3):389--405.

\bibitem[Ding et~al., 2019]{ding2019decomposing}
Ding, P., Feller, A., and Miratrix, L. (2019).
\newblock Decomposing treatment effect variation.
\newblock {\em Journal of the American Statistical Association}, 114(525):304--317.

\bibitem[Dudley, 2014]{dudley2014uniform}
Dudley, R.~M. (2014).
\newblock {\em Uniform central limit theorems}, volume 142.
\newblock Cambridge university press.

\bibitem[Fu et~al., 2020]{fu2020cvxr}
Fu, A., Narasimhan, B., and Boyd, S. (2020).
\newblock Cvxr: An r package for disciplined convex optimization.
\newblock {\em Journal of Statistical Software}, 94:1--34.

\bibitem[Gail and Simon, 1985]{gail1985testing}
Gail, M. and Simon, R. (1985).
\newblock Testing for qualitative interactions between treatment effects and patient subsets.
\newblock {\em Biometrics}, pages 361--372.

\bibitem[Hammer et~al., 1996]{hammer1996trial}
Hammer, S.~M., Katzenstein, D.~A., Hughes, M.~D., Gundacker, H., Schooley, R.~T., Haubrich, R.~H., Henry, W.~K., Lederman, M.~M., Phair, J.~P., Niu, M., et~al. (1996).
\newblock A trial comparing nucleoside monotherapy with combination therapy in hiv-infected adults with cd4 cell counts from 200 to 500 per cubic millimeter.
\newblock {\em New England Journal of Medicine}, 335(15):1081--1090.

\bibitem[Hsu, 2017]{hsu2017consistent}
Hsu, Y.-C. (2017).
\newblock Consistent tests for conditional treatment effects.
\newblock {\em The Econometrics Journal}, 20(1):1--22.

\bibitem[Hudson, 2023]{hudson2023nonparametric}
Hudson, A. (2023).
\newblock Nonparametric inference on non-negative dissimilarity measures at the boundary of the parameter space.
\newblock {\em arXiv preprint arXiv:2306.07492}.

\bibitem[Hudson et~al., 2026]{hudson2026inference}
Hudson, A., Carone, M., and Shojaie, A. (2026).
\newblock Inference on function-valued parameters using a restricted score test.
\newblock {\em Journal of the Royal Statistical Society Series B: Statistical Methodology}, page qkag043.

\bibitem[Kennedy, 2023]{kennedy2023towards}
Kennedy, E.~H. (2023).
\newblock Towards optimal doubly robust estimation of heterogeneous causal effects.
\newblock {\em Electronic Journal of Statistics}, 17(2):3008--3049.

\bibitem[Kennedy, 2024]{kennedy2024semiparametric}
Kennedy, E.~H. (2024).
\newblock Semiparametric doubly robust targeted double machine learning: a review.
\newblock {\em Handbook of statistical methods for precision medicine}, pages 207--236.

\bibitem[Kitagawa and Tetenov, 2018]{kitagawa2018should}
Kitagawa, T. and Tetenov, A. (2018).
\newblock Who should be treated? empirical welfare maximization methods for treatment choice.
\newblock {\em Econometrica}, 86(2):591--616.

\bibitem[Kosorok, 2008]{kosorok2008introduction}
Kosorok, M.~R. (2008).
\newblock {\em Introduction to empirical processes and semiparametric inference}, volume~61.
\newblock Springer.

\bibitem[Krause et~al., 2020]{krause2020covid}
Krause, P., Fleming, T.~R., Longini, I., Henao-Restrepo, A.~M., Peto, R., Dean, N., Halloran, M., Huang, Y., Fleming, T., Gilbert, P., et~al. (2020).
\newblock Covid-19 vaccine trials should seek worthwhile efficacy.
\newblock {\em The Lancet}, 396(10253):741--743.

\bibitem[Li et~al., 2024]{li2024estimation}
Li, Z., Nassif, H., and Luedtke, A. (2024).
\newblock Estimation of subsidiary performance metrics under optimal policies.
\newblock {\em arXiv preprint arXiv:2401.04265}.

\bibitem[Luedtke and Van Der~Laan, 2016]{luedtke2016statistical}
Luedtke, A.~R. and Van Der~Laan, M.~J. (2016).
\newblock Statistical inference for the mean outcome under a possibly non-unique optimal treatment strategy.
\newblock {\em Annals of statistics}, 44(2):713.

\bibitem[Piantadosi and Gail, 1993]{piantadosi1993comparison}
Piantadosi, S. and Gail, M. (1993).
\newblock A comparison of the power of two tests for qualitative interactions.
\newblock {\em Statistics in Medicine}, 12(13):1239--1248.

\bibitem[Rothman et~al., 1980]{rothman1980concepts}
Rothman, K.~J., Greenland, S., and Walker, A.~M. (1980).
\newblock Concepts of interaction.
\newblock {\em American journal of epidemiology}, 112(4):467--470.

\bibitem[Sanchez-Becerra, 2023]{sanchez2023robust}
Sanchez-Becerra, A. (2023).
\newblock Robust inference for the treatment effect variance in experiments using machine learning.
\newblock {\em arXiv preprint arXiv:2306.03363}.

\bibitem[Shi et~al., 2019]{shi2019sparse}
Shi, C., Lu, W., and Song, R. (2019).
\newblock A sparse random projection-based test for overall qualitative treatment effects.
\newblock {\em Journal of the American Statistical Association}.

\bibitem[Van~der Vaart, 2000]{van2000asymptotic}
Van~der Vaart, A.~W. (2000).
\newblock {\em Asymptotic statistics}, volume~3.
\newblock Cambridge university press.

\bibitem[van~der Vaart and Wellner, 1996]{vanderVaart1996}
van~der Vaart, A.~W. and Wellner, J.~A. (1996).
\newblock {\em Weak Convergence and Empirical Processes}.
\newblock Springer New York.

\bibitem[VanderWeele, 2009]{vanderweele2009distinction}
VanderWeele, T.~J. (2009).
\newblock On the distinction between interaction and effect modification.
\newblock {\em Epidemiology}, pages 863--871.

\bibitem[Westling, 2022]{westling2022nonparametric}
Westling, T. (2022).
\newblock Nonparametric tests of the causal null with nondiscrete exposures.
\newblock {\em Journal of the American Statistical Association}, 117(539):1551--1562.

\bibitem[Zelterman, 1990]{zelterman1990tests}
Zelterman, D. (1990).
\newblock On tests for qualitative interactions.
\newblock {\em Statistics \& probability letters}, 10(1):59--63.

\bibitem[Zhao et~al., 2012]{zhao2012estimating}
Zhao, Y., Zeng, D., Rush, A.~J., and Kosorok, M.~R. (2012).
\newblock Estimating individualized treatment rules using outcome weighted learning.
\newblock {\em Journal of the American Statistical Association}, 107(499):1106--1118.

\end{thebibliography}

\appendix
\newpage 

\numberwithin{equation}{section}

\section{Explanation of estimands as policy evaluation metrics}\label{appendix:policy}
First,
\begin{align*}
E_0\{Y(g_{f^*})\}=E_0[\{Y(1)-Y(0)\}f^*(X_s)]+E_0\{Y(0)\}.
\end{align*}
Then, from the definitions of minima and maxima,
\begin{align*}
\theta^{\text{min}}_{0,0}(f^*)  &=   \frac{1}{2}\{\theta^+_{0,0}(f^*)-\theta^-_{0,0}(f^*)\}-\frac{1}{2}|\theta^+_{0,0}(f^*)+\theta^-_{0,0}(f^*)|\\
&= E_0[\{Y(1)-Y(0)\}f^*(X_s)]-\frac{1}{2}\{\tau_0+|\tau_0|\}\\
&= E_0[\{Y(1)-Y(0)\}f^*(X_s)]-\frac{1}{2}[E_0\{Y(1)\}-E_0\{Y(0)\}+|\tau_0|]
\end{align*}
and 
\begin{align*}
\max [ E_0\{Y(0)\}, E_0\{Y(1)\}]&=\frac{1}{2}[E_0\{Y(1)\}+E_0\{Y(0)\}]+\frac{1}{2}|\tau_0|
\end{align*}
so
\begin{align*}
\theta^{\text{min}}_{0,0}(f^*)+ \max [ E_0\{Y(0)\}, E_0\{Y(1)\}]&=E_0[\{Y(1)-Y(0)\}f^*(X_s)]+E_0\{Y(0)\}\\&=
E_0\{Y(g_{f^*})\}.
\end{align*}

Moreover, 
\begin{align*}
\theta^+_{0,\tau_0}(f^*)&=E_0[\{Y(1)-Y(0)-\tau_0\}f^*(X_S)]\\
&=E_0\{Y(g_{f^*})\} -\tau_0E_0\{f^*(X_s)\}-E_0\{Y(0)\}\\
&=E_0\{Y(g_{f^*})\}-E_0\{Y(1)\}E_0\{f^*(X_s)\}-E_0\{Y(0)\}E_0\{1-f^*(X_s)\}
\end{align*}
where the second equality is obtained by adding and subtracting $E_0\{Y(0)\}$.

\section{Additional details on procedure}\label{appendix:bonus}

\noindent \textbf{Approach 2*: Basis Expansion with Structure Constraint}

Let $\mathcal{H} = h_1 \oplus h_2 \oplus \cdots $ be a vector space defined as the span of basis vectors $h_1, h_2, \ldots$ from $\mathbb{R}^{|s|}$ to $\mathbb{R}$.
 Let $J: \mathcal{H} \to \mathbb{R}^+$ be a measure of complexity for any $h \in \mathcal{H}$. We set $\mathcal{F}$ as
 \begin{align*}
     \mathcal{F}:= \left\{f: x_s \mapsto \kappa^{-1}\left(b_0 + \sum_{k=1}^\infty b_k h_k(x_s)\right): b_0, b_1, b_2, \ldots \in \mathbb{R}, 
     J\left( \sum_{k=1}^\infty b_k h_k \right) \leq \lambda \right\}
 \end{align*}
 for some $\lambda > 0$ and link function $\kappa$.
 The constraint $J\left( \sum_{k=1}^\infty b_k h_k \right) \leq \lambda$ enforces an upper bound on the smoothness of any function $f$ and is selected so that the requisite Donsker conditions hold. For the purpose of identifiability, we also assume that the basis functions $h_1, h_2, \ldots$ are centered to have zero mean. Approach 2 in the main manuscript is a special case of the above. 

In practice, it may not be obvious to the analyst how to choose the tuning parameter $\lambda$. Theorem 2 of \citet{hudson2026inference} suggests that $\lambda$ can be selected data-adaptively without compromising type I error, so long as the estimated choice of $\mathcal{F}$ converges to a fixed class. However, empirical results suggest that type I error inflation can occur with data-driven tuning parameter selection. Furthermore, certain choices of link function $\kappa$ and constraint $\lambda$ may lead to the corresponding optimization problem being non-convex and hence difficult to solve. Closely related optimization problems are considered for learning optimal treatment regimes \citep{zhao2012estimating,athey2021policy}, where a surrogate objective is often used due to challenges in implementation.

\subsection{Extension to variance-weighted estimands}\label{appendix:var_weight}
Conceptually, we would expect that detecting heterogeneity should be easier when the CATE sees a greater departure from the ATE (for quantitative heterogeneity) or $\delta$ (for qualitative heterogeneity) for values of $X$ for which heterogeneity can be measured with greater precision. Thus, with the aim of increasing power, we  describe a slight modification of our procedure based on variance weighting.

Let $V_{0,\tau_0}(f)$, $V^+_{0,
\delta}(f)$, and $V^-_{0, \delta}(f)$ denote the variance of the efficient influence functions of $\theta^+_{0,\tau_0} - \theta^{-}_{0,\tau_0}$, $\theta^+_{0, \delta}$, and $\theta^{-}_{0,\delta}$, respectively:
\begin{align*}
    &V_{0, \tau_0}(f) = E_0\left[ \left\{\varphi^+_{P_0, \tau_0}(Z;f) - \varphi^{-}_{P_0, \tau_0}(Z;f) \right\}^2\right],
    \\ 
    &V^+_{0, \delta}(f) = E_0\left[ \left\{\varphi^+_{P_0, \delta}(Z;f) \right\}^2\right]
    ,\quad 
    V^{-}_{0, \delta}(f) = E_0\left[ \left\{\varphi^{-}_{P_0, \delta}(Z;f) \right\}^2\right].
\end{align*}
Suppose that each of $V_{0,\tau_0}$, $V^+_{0,
\delta}$, and $V^-_{0, \delta}$ is bounded away from zero uniformly in $\mathcal{F}$ and that uniformly consistent estimators $V_{n, \tau_n}$, $V^+_{n, \delta}$ and $V^{-}_{n,\delta}$ are available.
For instance, a natural approach to estimator construction is to use the sample average of plug-in estimators for the efficient influence functions. We suspect that one could perform more powerful tests for heterogeneity using a similar approach as described above, simply replacing $\sup_{f \in \mathcal{F}}|\theta^{+}_{n,\tau_n}(f) - \theta^{-}_{n,\tau_n}(f)|$ (for the quantitative test) and $\sup_{f \in \mathcal{F}}\theta^+_{n,\delta}(f)$ and $\inf_{f \in \mathcal{F}}\theta^-_{n,\delta}(f)$ (for the qualitative test) with their variance-weighted counterparts
\begin{align*}
    &\sup_{f \in \mathcal{F}}\left|V^{-1/2}_{n,\tau_n}(f)\left\{\theta^{+}_{n,\tau_n}(f) - \theta^{-}_{n,\tau_n}(f) \right\}\right|,
    \\
    &\sup_{f \in \mathcal{F}} \left\{V^+_{n,\delta}(f)\right\}^{-1/2}\theta^+_{n,\delta}(f), 
    \quad
    \inf_{f \in \mathcal{F}} \left\{V^-_{n,\delta}(f)\right\}^{-1/2}\theta^-_{n,\delta}(f).
\end{align*}
As for the unweighted approach, the multiplier bootstrap may be used to approximate the relevant null distributions.
Moreover, analogous type I error control and power results can be readily established through an application of Slutsky's theorem. While we expect the variance-weighted approach to outperform the unweighted method in some instances, we reserve a formal theoretical comparison of their power for future work.

\section{Additional simulation design details and results} \label{appendix:simulations}

\subsection{Hypothesis testing approaches under consideration} \label{appendix:simulations-methods}

In what follows, we first briefly describe the kernel ridge regression and sample splitting tests considered in our simulation study.
We then present an approach for implementing our proposal using Approach 1 (linear threshold rules) with multivariable $X_s$.
\\
\textbf{Kernel ridge regression}
\\
We first obtained estimates for the conditional mean $\mu_{0}$ and propensity score $\pi_0$ using the highly adaptive LASSO, allowing us to construct $\psi_n$.
We then obtain a DR-learner estimate of the CATE $\tau_{0,s}$ by estimating the conditional mean of $\psi_n(Z)$ given $X_s$ using kernel ridge regression.

To be explicit, suppose each element of $X_s$ is scaled to have unit variance.
Let $\mathbf{K}$ be an $n \times n$, matrix with $K_{ij} = \exp\left(-\|X_{s,i} - X_{s,j} \|_2^2 \right)$.
Let the eigenvectors and eigenvalues, respectively, be given by $\kappa_1, \kappa_2, \ldots, \kappa_n$ and $\iota_1, \iota_2, \ldots, \iota_n$.
For simplicity we allow the eigenbasis to be truncated at a large level $D$ (taken to be $n/2$ in our simulations).
Let $\boldsymbol{\kappa} = (\kappa_1, \kappa_2, \cdots, \kappa_D)$, let $\mathcal{I} = \text{diag}(\iota_1, \iota_2, \ldots, \iota_n)$, and $\boldsymbol{\psi}_n = (\psi_n(Z_1), \psi_n(Z_2), \ldots, \psi_n(Z_n) )^\top$
We estimate the CATE as
\begin{align*}
    \tau^\lambda_{n,s}(X_{s,i}) = (\kappa_{1,i}, \kappa_{2,i}, \ldots ,\kappa_{D,i})\left(\boldsymbol{\kappa}^\top
    \boldsymbol{\kappa} + \lambda \mathcal{I} \right)^{-1} \boldsymbol{\kappa}^\top \boldsymbol{\psi}_n,
\end{align*}
where $\lambda > 0$ is a tuning parameter that is selected using cross-validation.
Let the evaluation of the CATE estimator on the observed $X_s$ be denoted by \\$\boldsymbol{\tau}_{n,s}^\lambda = \left(\tau^\lambda_{n,s}(X_{s,1}), \tau^\lambda_{n,s}(X_{s,2}), \ldots, \tau^\lambda_{n,s}(X_{s,n}) \right)^\top$.

We estimate $\theta^+_\tau - \theta^-_\tau$ as
\begin{align*}
    T_n =
    \left\|\left(\boldsymbol{\kappa}^\top \boldsymbol{\kappa} + \lambda \mathcal{I} \right)^{-1} \boldsymbol{\kappa}^\top \boldsymbol{\psi}_n
     \left(I - n^{-1}\mathbf{1}\mathbf{1}^\top \right) \right\|_{1}.
\end{align*}
We approximate the limiting distribution of $T_n$ using Monte Carlo sampling.
We simulate random vectors $B^*$ from a multivariate normal distribution with mean zero and covariance
\begin{align*}
    \Sigma = \left(\boldsymbol{\kappa}^\top \boldsymbol{\kappa} + \lambda \mathcal{I} \right)^{-1} \boldsymbol{\kappa}^\top \left(\boldsymbol{\psi}_n - \boldsymbol{\tau}_{n,s}^\lambda\right)
     \left(I - n^{-1}\mathbf{1}\mathbf{1}^\top \right) 
     \left(\boldsymbol{\psi}_n - \boldsymbol{\tau}_{n,s}^\lambda\right)^\top \boldsymbol{\kappa} \left(\boldsymbol{\kappa}^\top \boldsymbol{\kappa} + \lambda \mathcal{I} \right)^{-1}.
\end{align*}
We then obtain an asymptotic approximation of the $p$-value as the sample proportion of Monte Carlo samples $\|B^*\|_1$ that exceed $T^*$.

We estimate $\theta^+_\delta$ and $\theta^-_{\delta}$ as
\begin{align*}
    T^+_n =\left\|\left(\boldsymbol{\kappa}^\top \boldsymbol{\kappa} + \lambda \mathcal{I} \right)^{-1} \boldsymbol{\kappa}^\top \boldsymbol{\psi}_n
     \text{diag}(\boldsymbol{\tau}_{n,s}^\lambda > \delta) \right\|_{1},
     \\
     T^-_n =\left\|\left(\boldsymbol{\kappa}^\top \boldsymbol{\kappa} + \lambda \mathcal{I} \right)^{-1} \boldsymbol{\kappa}^\top \boldsymbol{\psi}_n
     \text{diag}(\boldsymbol{\tau}_{n,s}^\lambda > \delta) \right\|_{1}.
\end{align*}
Similarly as for the qualitative test, we use a Monte Carlo sampling approach to determine when to reject the null of no qualitative heterogeneity.
We draw random samples $C^*$ from a multivariate normal distribution with mean zero and covariance
\begin{align*}
    \Sigma = \left(\boldsymbol{\kappa}^\top \boldsymbol{\kappa} + \lambda \mathcal{I} \right)^{-1} \boldsymbol{\kappa}^\top \left(\boldsymbol{\psi}_n - \boldsymbol{\tau}_{n,s}^\lambda\right)
     \left(\boldsymbol{\psi}_n - \boldsymbol{\tau}_{n,s}^\lambda\right)^\top \boldsymbol{\kappa} \left(\boldsymbol{\kappa}^\top \boldsymbol{\kappa} + \lambda \mathcal{I} \right)^{-1}.
\end{align*}
We reject the null when each of $T_n^+$ and $T_n^-$ exceed the $1 - \alpha$ quantiles of $\|C^*\text{diag}(C^* > \delta)\|_1$ and $\|C^*\text{diag}(C^* < \delta)\|_1$.
\\
\textbf{Sample splitting}
\\
First the data is split into two partitions.
Let $\Delta$ be a binary variable that indicates membership of an observation in the first partition.
On the first partition of the data ($\Delta = 1$), we obtain a DR-learning estimate for the optimal treatment rule as follows.
Using the method described in the preceding subsection, we obtain a kernel ridge regression estimate for the CATE, $\tau^\Delta_{n,s}$.
For fixed $\delta$, we estimate the optimal rule as
\begin{align*}
    g^\Delta_{n, \delta}: x_s \mapsto \mathds{1}\left(\tau^\Delta_{n,s}(x_s) > \delta\right),
\end{align*}
and with $\delta = \tau_0$, we estimate the optimal rule as
\begin{align*}
    g^\Delta_{n, \tau}: x_s \mapsto \mathds{1}\left(\tau_{n,s}(x_s) > -\frac{1}{\sum \Delta_i} \sum \tau^\Delta_{n,s} \Delta_i\right).
\end{align*}
The second partition ($\Delta = 0$) of the data is then used to perform a test of heterogeneity using our proposed method with the singleton set $\mathcal{F} = \{g^\Delta_{n,\tau}\}$ for quantitative heterogeneity and $\mathcal{F} = \{g^\Delta_{n, \delta}\}$ for qualitative heterogeneity.
\\ 
\textbf{Linear threshold rules for multidimensional $X_s$}
\\
We now develop an implementation of our proposal when $\mathcal{F}$ is a linear threshold class (Approach 1).
We approximate $\mathcal{F}$ using a discrete class so that all optimizations can be performed by evaluating the maximum or minimum over a finite countable set.

We first scale $X_s$ so that it belongs to the unit cube $[0,1]^{|s|}$.
Then, for a positive integer $k_1$, we generate $k_1^{|s| - 1}$ evenly-spaced grid points $\gamma^1, \ldots, \gamma^{k_1^{|s| - 1}}$ on the cube $[0, \pi/2]^{|s| - 1}$.
For any $j$ between $1$ and $k_1^{|s| - 1}$, we define $\omega^j$ as
\begin{align*}
    &\omega^j_1 = \cos(\gamma^j_1),
    \\
    &\omega^j_2 = \sin(\gamma^j_1) \cos(\gamma^j_2)
    \\
    &\omega^j_3 = \sin(\gamma^j_1) \sin (\gamma^j_2) \cos(\gamma^j_3)
    \\
    &\vdots
    \\
    &\omega^j_{|s| - 1}  = \sin(\gamma^j_1) \cdots \sin (\gamma^j_{|s| - 2}) \cos (\gamma^j_{|s| - 1})
    \\
    &\omega^j_{|s|}= \sin(\gamma^j_1) \cdots \sin (\gamma^j_{|s| - 2}) \sin (\gamma^j_{|s| - 1}), 
\end{align*}
and let $u^j = (\sum_i \omega^j_i)^{-1}\omega^j$.
Now, for a positive integer $k_2$, let $v_1, v_2, \ldots, v_{k_2}$ be a set of evenly spaced grid points on the interval $[0,1]$.
We finally set $\mathcal{F}$ as
\begin{align*}
    \tilde{\mathcal{F}} := &\left\{x_s \mapsto \mathds{1}((u^{j_1})^\top x_s \geq v_{j_2} ): j_1 = 1, \ldots, k^{|s| - 1}, j_2 = 1, \ldots, k_2 \right\} \cup
    \\
    &\left\{x_s \mapsto \mathds{1}((u^{j_1})^\top x_s \leq v_{j_2} ): j_1 = 1, \ldots, k^{|s| - 1}, j_2 = 1, \ldots, k_2 \right\} .
\end{align*}

\subsection{Simulation design for multivariable effect modifier}\label{appendix:simulations-design}

Let $X = (X_1, X_2, \ldots,  X_5)$ be a vector of independent uniform random variables on the interval $[-1, 1]$.
We generate the treatment assignment variable $A$ from the conditional distribution $\pi_0(1|x) = \text{expit}\left\{\frac{1}{8}x_1 + \frac{1}{4}\sin(\pi x_2) \right\}$. Given $X$ and $A$, we generate the outcome as $Y = h(X_1, X_2, X_3,X_4,X_5) + A \gamma(X_3) + \epsilon$,
where $h(x) = \frac{1}{2}\left\{x_1 + x_4 + \text{sign}(x_5)x_5^2 \right\} + \text{expit}\left\{\frac{1}{2}\left(x_2 + x_3\right)\right\}$, $\epsilon \sim N(0, 3^2)$ is white noise, and  $\gamma$ is a function of $X_3$ that we manipulate.
It can be seen that the CATE, given $X_3$, is equal to $\gamma$.
Our simulations assess various tests' performance for simultaneously assessing heterogeneity in $(X_1, X_2, X_3)$.

We consider the following specifications of $\gamma$ to control whether the qualitative and/or qualitative nulls hold:
\begin{itemize}
    \item Setting 1 (No heterogeneity)
    $\gamma(x_1, x_2, x_3) = \frac{1}{4}$.
    \item \sloppy Setting 2 (Quantitative heterogeneity; monotone CATE): $\gamma(x_1, x_2, x_3) = 5\left(x_1 + x_2 + x_3 - 1\right)\mathds{1}(x_1 + x_2 + x_3 >1)$.
    \item Setting 3 (Qualitative heterogeneity; monotone CATE): $\gamma(x_1, x_2, x_3) = \frac{1}{2}\text{sign}(x_1 + x_2 + x_3) (x_1 + x_2 + x_3)^2$.
\end{itemize}
We note that under Settings 2 and 3, the optimal rule belongs to the class of linear threshold functions (Approach 1).
We consider the following methods (abbreviated names in Figures \ref{fig:sims-quant-null-multi}-\ref{fig:sims-qual-alt-multi} are in quotes):
\begin{itemize}
\item Quant (Qual) Linear Threshold: our proposed quantitative (qualitative) test implemented using linear threshold rules, i.e., Approach 1 in Section \ref{sec:imp}. The implementation described above in Section \ref{appendix:simulations-methods} is used with $k_1 = k_2 = 10$ (`Linear Threshold').
\item Quant (Qual) Kernel ridge: test based on kernel ridge regression DR-learning estimator described in Section \ref{appendix:simulations-methods} (`Ridge'). 
\item Quant (Qual) Sample-split: sample splitting test described in Section \ref{appendix:simulations-methods} (`Split').
\end{itemize}

As in the main text, the conditional mean and propensity score are estimated using the highly adaptive lasso \citep{benkeser2016highly}.  All tests were performed at the nominal level $\alpha = .05$. Under each of the above settings, we generated 1000 random data sets for $n \in \{250, 500, 1000, 2000\}$.

\subsection{Simulation results for multivariable effect modifier}

\begin{figure}[!h]
    \centering
    \includegraphics[width=0.75\linewidth]{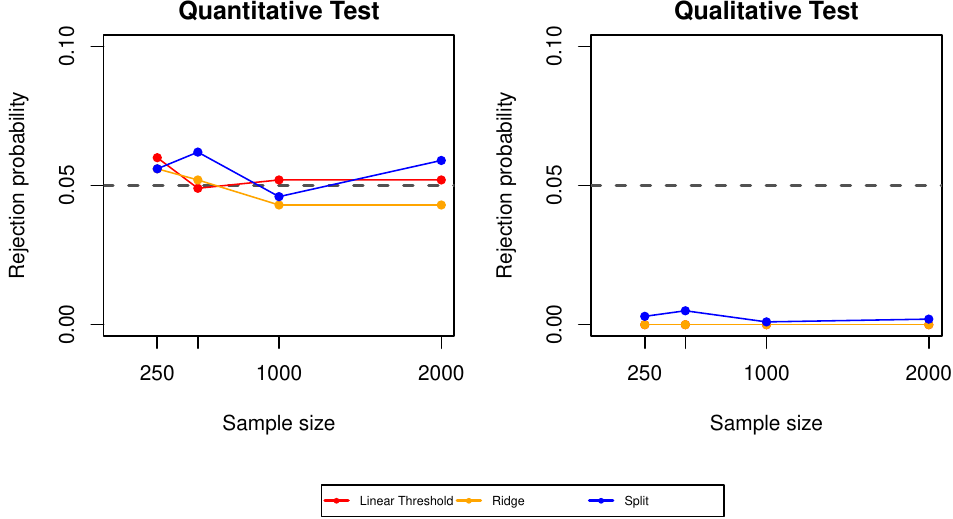}
    \caption{Monte Carlo estimate of rejection probability, when the null hypothesis of no quantitative or qualitative qualitative heterogeneity holds. The definitions of the abbreviations in the legend are given in Appendix \ref{appendix:simulations-design}} 
\label{fig:sims-quant-null-multi}
\end{figure}

\begin{figure}[!h]
    \centering
    \includegraphics[width=0.75\linewidth]{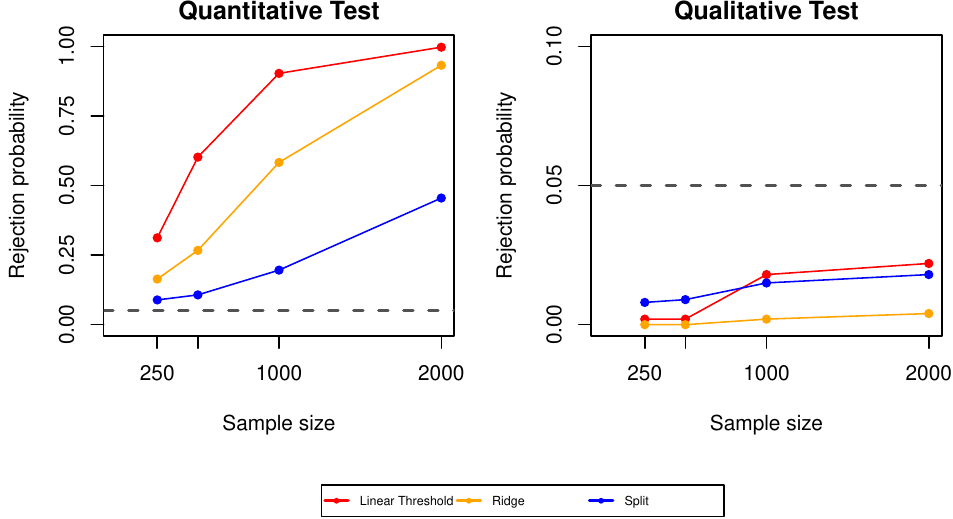}
    \caption{Monte Carlo estimate of rejection probability in the presence of quantitative heterogeneity and the absence of qualitative heterogeneity.} 
\label{fig:sims-quant-alt-multi}
\end{figure}

\begin{figure}[!h]
    \centering
    \includegraphics[width=0.75\linewidth]{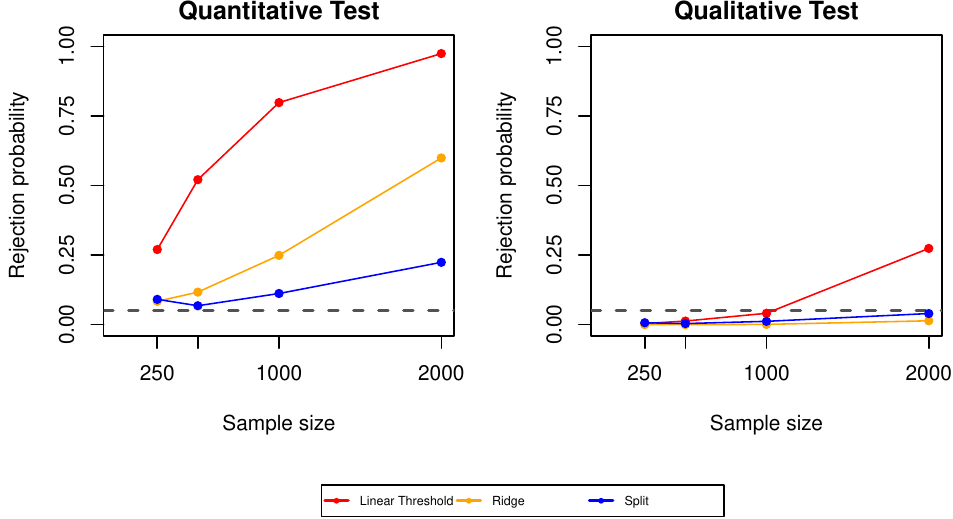}
    \caption{Monte Carlo estimate of rejection probability in the presence of qualitative heterogeneity.} 
\label{fig:sims-qual-alt-multi}
\end{figure}

Figure \ref{fig:sims-quant-null-multi} displays rejection rates under the null of no heterogeneity. In this setting, all tests of the quantitative null have rejection rates near the nominal level for all $n$.
Similarly as seen when assessing the univariable tests, the multivariable test for qualitative heterogeneity has rejection rate far lower than 0.05.
Rejection rates in the presence of quantitative heterogeneity and absence of qualitative heterogeneity are shown in Figure \ref{fig:sims-quant-alt-multi}.
Among all quantitative tests, the quantitative linear threshold rule method performs best.
Notably, the sample splitting test has relatively poor power even in large samples.
All qualitative tests reject at a rate lower than $\alpha$, as expected.
Figure \ref{fig:sims-qual-alt-multi} shows rejection rates in the presence of qualitative heterogeneity.
Quantitative tests perform similarly as when there is quantitative but not qualitative effect heterogeneity.
Among methods qualitative testing, only the linear threshold approach yields power exceeding $\alpha$ at $n = 2000$, and power is modest even in this case.
This suggests that the in the multivariable setting, qualitative heterogeneity remains difficult to detect in the absence of large signal.

\section{Asymptotic behaviour under fixed null and alternatives}\label{appendix:fixed}

We begin by establishing type I error control for both tests. Recall from Section \ref{sec:construct} that our test for quantitative heterogeneity rejects when $n^{1/2}\sup_{f \in \mathcal{F}}|\theta^+_{n,\tau_n}(f) - \theta^-_{n,\tau_n}(f)|$ exceeds the $(1-\alpha)$ quantile of its null limiting distribution. 
The next result, which follows immediately from Corollary \ref{corollary:weak}, states that our test controls the type I error level.
\begin{theorem}\label{theorem:size-quant}
(Asymptotic type I error control : quantitative heterogeneity) 
Let $P_0$ be any probability distribution for which  $ H_0^{\mathrm{I},*}$ holds.
Let $t_\alpha$ be the $(1-\alpha)$ quantile of $\sup_{f \in \mathcal{F}}|\mathbb{G}(f)|$, where $\mathbb{G}$ is defined in Corollary \ref{corollary:weak}, and we assume the distribution function of $\sup_{f \in \mathcal{F}}|\mathbb{G}(f)|$ is continuous in a neighborhood of $t_\alpha$. Then under the conditions of Theorem \ref{theorem:al},
\begin{align*}
    \lim_{n \to \infty}P_0\left(n^{1/2}\sup_{f \in \mathcal{F}}|\theta^+_{n,\tau_n}(f) - \theta^-_{n,\tau_n}(f)| > t_\alpha\right) = \alpha.
\end{align*}
\end{theorem}

Establishing type I error control for the qualitative test is more involved because, as discussed in Section \ref{sec:qual-estimands}, $H_0^{\mathrm{II}}$ is a composite null hypothesis.
We are therefore required to show that if $\{\theta_{0,\delta}^+(f):f \in \mathcal{F}\}$ and $\{\theta^-_{0,\delta}(f): f \in \mathcal{F}\}$ take any value compatible with the null, the probability of rejecting the null does not exceed $\alpha$.
In what follows, we argue that the procedure proposed in Section \ref{sec:construct} yields type I error control.

\begin{theorem}\label{theorem:size-qual}(Asymptotic type I error control: qualitative heterogeneity)
Suppose $P_0$ is any fixed probability distribution for which $ H_0^{\mathrm{II},*}$ holds. 
Let $t^+_\alpha$ and $t^-_\alpha$ be chosen as the $(1-\alpha)$ quantile of $\sup_{f \in \mathcal{F}} \mathbb{G}^+(f)$ and the $\alpha$ quantile of $\inf_{f \in \mathcal{F}} \mathbb{G}^-(f)$ respectively (see Corollary \ref{corollary:weak}); the limiting distribution functions are assumed to be continuous. Then under the conditions of Theorem \ref{theorem:al},
\begin{align*}
    \limsup_{n\to \infty} P_0\left( n^{1/2}\sup_{f \in \mathcal{F}}\theta_{n,\delta}(f) > t^{+}_{\alpha}
    \text{ and }
    n^{1/2}\inf_{f \in \mathcal{F}} \theta_{n,\delta}(f) < t^{-}_{\alpha}\right)\leq \alpha.
\end{align*}
\end{theorem}

The following theorems establish consistency for both tests.

\begin{theorem}\label{theorem:power_f-quant}
(Power against fixed alternatives: quantitative heterogeneity) Let $P_0$ be any distribution for which $\sup_{f\in \mathcal{F}}|\theta^+_{0,\tau_0}(f) - \theta^-_{0,\tau_0}(f)| > 0$.
Then under the conditions of Theorem \ref{theorem:al}, and assuming  that the distribution function of $\sup_{f \in \mathcal{F}}|\mathbb{G}(f)|$ is continuous,
\begin{align*}
    \lim_{n \to \infty}P_0\left(n^{1/2}\sup_{f \in \mathcal{F}}|\theta^+_{n,\tau_n}(f) - \theta^-_{n,\tau_n}(f)| > t_\alpha\right) = 1.
\end{align*}
\end{theorem}
\begin{theorem}\label{theorem:power_f-qual}
    (Power against fixed alternatives for qualitative heterogeneity) Let $P_0$ be any distribution for which $\sup_{f \in \mathcal{F}} \theta^+_{0,\delta}(f) > 0$ and $\inf_{f \in \mathcal{F}} \theta^-_{0, \delta}(f) < 0$. Then under the conditions of Theorem \ref{theorem:al}, and assuming the distribution functions of $\sup_{f \in \mathcal{F}} \mathbb{G}^+(f)$ and $\inf_{f \in \mathcal{F}} \mathbb{G}^-(f)$ are continuous,
    \begin{align*}
    \lim_{n \to \infty}
    P_0\left(n^{1/2} \sup_{f \in \mathcal{F}}\theta_{n,\delta}(f) > t^{+}_{\alpha}
    \text{ and }
    n^{1/2}\inf_{f \in \mathcal{F}} \theta_{n,\delta}(f) < t^{-}_{\alpha}\right) = 1.
\end{align*}
\end{theorem}

\section{Local Asymptotic Behavior}\label{appendix:local}

\subsection{Test for quantitative heterogeneity}

In what follows, we will investigate the properties of our tests in a local asymptotic framework. We will consider first quantitative and then qualitative heterogeneity testing. The first case follows along the lines e.g. of Section 3.10 of  \citet{vanderVaart1996} and \citet{westling2022nonparametric}. We will devote more attention to second case given the complexities that arise due to the null hypothesis being composite.

Let $P_0$ be a probability distribution for which $\sup_{f\in \mathcal{F}}|\theta^{+}_{0,\tau_0}(f)-\theta^{-}_{0,\tau_0}(f)|=0$.  We define $S: \mathcal{Z} \to \mathbb{R}$ as a score function with mean zero and finite variance under $P_0$, where $\mathcal{Z}$ is the sample space corresponding to $P_0$. Also,
\[c : f \mapsto \int S(z)\{\varphi^+_{0,\tau_0}(z;f)-\varphi^-_{0,\tau_0}(z;f)\}dP_0(z).\]
We will consider local alternative distributions $P_n$ that satisfy
\begin{align}\label{eq:local_alt}
 \lim_{n\to \infty}\int \left[ n^{1/2}\{dP_n(z)^{1/2}-dP_0(z)^{1/2}\}-\frac{1}{2}S(z)dP_0(z)^{1/2}\right]^2=0.
 \end{align}

 \begin{theorem}\label{theorem:local_quant}(Weak convergence under local data generating laws: quantitative heterogeneity)
 Suppose that our data are drawn as an i.i.d. triangular array $Z_{n,1},...,Z_{n,n}$ from some sequence $P_n$ in \eqref{eq:local_alt}, and that $\sup_{f\in \mathcal{F}}|c(f)|$ is bounded. Then under the conditions of Theorem \ref{theorem:al},
 \[\theta^+_{n,\tau_n}(f)-\theta^-_{n,\tau_n}(f)=\frac{1}{n}\sum^n_{i=1}\{\varphi^+_{0,\tau_0}(Z_{n,i};f)-\varphi^-_{0,\tau_0}(Z_{n,i};f)\}+r_n(f)\]
 where $\sup_{f\in \mathcal{F}}|n^{1/2}r_n(f)|$ converges to zero in probability under sampling from $P_n$. Moreover, $\{n^{1/2}\{\theta^+_{n,\tau_n}(f)-\theta^-_{n,\tau_n}(f)\}:f\in \mathcal{F}\}$ converges weakly under $P_n$ to $\{\mathbb{G}(f)+c(f):f\in \mathcal{F}\}$ as an element of $l^\infty(\mathcal{F})$. Here, $\mathbb{G}$ is a tight, mean-zero Gaussian process with covariance $\Sigma: (f_1, f_2) \mapsto P_0[\{\varphi^+_{0,\tau_0}(f_1)-\varphi^-_{0,\tau_0}(f_1)\}\{\varphi^+_{0,\tau_0}(f_2) - \varphi^-_{0,\tau_0}(f_2)\}]$.
  \end{theorem}

 This can then be converted into a result on power under local alternatives.
 \begin{corollary}\label{corollary:local_quant}(Power under local alternatives: quantitative heterogeneity)
Let $t_{\alpha}$  be the $(1-\alpha)$ quantile of $\sup_{f \in \mathcal{F}}|\mathbb{G}(f)|$. Then under sampling from $P_n$,
and the conditions of Theorems \ref{theorem:al} and \ref{theorem:local_quant},
 \[\lim_{n \to \infty} P_n \left(n^{1/2}\sup_{f\in \mathcal{F}}|\theta^+_{n,\tau_n}(f)-\theta^-_{n,\tau_n}(f)|>t_\alpha\right)>\alpha.\]
 \end{corollary}

 Hence our test has power to detect alternatives that shrink to the null at the $n^{-1/2}$-rate, which is the same rate as in parametric testing problems. 

\subsection{Test for qualitative heterogeneity}
\subsubsection{Type I error control}

In what follows, we show that the type I error of our procedure is preserved in the following two instances:
\begin{enumerate}
    \item $\sup_{f \in \mathcal{F}}\theta^+_{0,\delta}(f) \downarrow 0$, and $\inf_{f \in \mathcal{F}}\theta^{-}_{0,\delta}(f) = 0$.
    \item $\sup_{f \in \mathcal{F}}\theta^+_{0,\delta}(f) = 0$, and $\inf_{f \in \mathcal{F}}\theta^{-}_{0,\delta}(f) \uparrow 0$.
\end{enumerate}
To accomplish the first instance, suppose $P_0$ is a probability distribution for which $\theta^+_{0,\delta}(f)=\theta^-_{0,\delta}(f)=0$  $\forall f\in \mathcal{F}$ and hence $\sup_{f \in \mathcal{F}}\theta^{+}_{0,\delta}(f) = \inf_{f\in\mathcal{F}}\theta^{-}_{0,\delta}(f) = 0$. 
We respectively define $S^+:\mathcal{Z}\to \mathbb{R}$ and $S^-:\mathcal{Z} \to \mathbb{R}$ as score functions with mean zero under $P_0$, which also satisfy
\begin{align*}
    \sup_{f \in \mathcal{F}} \int S^+(z) \varphi^+_{0,\delta}(z;f)dP_0(z) > 0,  \quad 
    \inf_{f \in \mathcal{F}} \int S^+(z) \varphi^-_{0,\delta}(z;f)dP_0(z) \geq 0, 
    \\
     \sup_{f \in \mathcal{F}} \int S^-(z) \varphi^+_{0,\delta}(z;f)dP_0(z) \leq 0,  \quad 
    \inf_{f \in \mathcal{F}} \int S^-(z) \varphi^-_{0,\delta}(z;f)dP_0(z) < 0. 
\end{align*}
We define the $P^+_n$ and $P^-_n$ as sequences of probability distributions that approach $P_0$ from the paths $S^+$ and $S^-$, respectively, in the sense that
\begin{align*}
&\lim_{n\to \infty}\int \left[ n^{1/2}\{dP^+_n(z)^{1/2}-dP_0(z)^{1/2}\}-\frac{1}{2}S^+(z)dP_0(z)^{1/2}\right]^2 = 0 
\\
&\lim_{n\to \infty}\int \left[ n^{1/2}\{dP^-_n(z)^{1/2}-dP_0(z)^{1/2}\}-\frac{1}{2}S^-(z)dP_0(z)^{1/2}\right]^2 = 0.
\end{align*}
To aid interpretation, the following general lemma shows that the local laws $P^+_n$ and $P^-_n$ are compatible with scenarios 1 and 2. 

\begin{lemma}\label{lemmma:local}
Let $P_0$ be a distribution for which $\theta^+_{0,\delta}(f)=\theta^-_{0,\delta}(f)=0$ $\forall f\in \mathcal{F}$ and therefore $\sup_{f \in \mathcal{F}} \theta^+_{0,\delta}(f) = \inf_{f \in \mathcal{F}}\theta^-_{0,\delta}(f) = 0.$
Let $S:\mathcal{Z} \to \mathbb{R}$ be a score function with zero mean and finite variance, and define $c^+:\mathcal{F} \to \mathbb{R}$ and $c^-:\mathcal{F} \to \mathbb{R}$ as
\begin{align*}
    &c^+: f \mapsto \int S(z) \varphi^+_{0,\delta}(z;f)dP_0(z),
    \\
    &c^-: f \mapsto 
\int S(z) \varphi^-_{0,\delta}(z;f)dP_0(z).
\end{align*}
Suppose that $\sup_{f \in \mathcal{F}}|c^+(f)|$ and $\sup_{f \in \mathcal{F}}|c^-(f)|$ are bounded and consider a sequence of distributions $P_n$ which satisfy
\begin{align}
    &\lim_{n\to \infty}\int \left[ n^{1/2}\{dP_n(z)^{1/2}-dP_0(z)^{1/2}\}-\frac{1}{2}S(z)dP_0(z)^{1/2}\right]^2 = 0,
    \label{eq:local_alt_general}
\end{align}
and moreover that $\sup_{f\in \mathcal{F}}|R^+(P_n,P_0)|=o(n^{-1/2})$ and $\sup_{f\in \mathcal{F}}|R^-(P_n,P_0)|=o(n^{-1/2})$, where
\begin{align*}
R^+(P_n,P_0)&\equiv \theta^+_{P_n,\delta}(f)-\int \varphi_{0,\delta}^+(z;f)\{dP_n(z)-dP_0(z)\},\\
R^-(P_n,P_0)&\equiv \theta^-_{P_n,\delta}(f)-\int \varphi_{0,\delta}^-(z;f)\{dP_n(z)-dP_0(z)\}.
\end{align*}
Then it follows that 
\begin{align*}
&\lim_{n \to \infty}\sup_{f\in \mathcal{F}} \theta^{+}_{P_n,\delta}(f)=0, \quad \quad \lim_{n \to \infty}\inf_{f \in \mathcal{F}}  \theta^{-}_{P_n,\delta}(f)=0
\end{align*}
and moreover
\begin{align}
    &\lim_{n \to \infty} \sup_{f\in \mathcal{F}}n^{1/2}\theta^{+}_{P_n,\delta}(f) = \sup_{f \in \mathcal{F}} c^+(f),\quad \quad 
    \lim_{n \to \infty}\inf_{f \in \mathcal{F}} n^{1/2}\theta^{-}_{P_n,\delta}(f) =\inf_{f \in \mathcal{F}} c^-(f).\label{eq:lim_score}
    \end{align}
\end{lemma}
By application of the above, it follows immediately that 
\begin{align*}
    &\lim_{n \to \infty} \sup_{f\in \mathcal{F}} n^{1/2}\theta^{+}_{P^+_n,\delta}(f) > 0, \quad \quad 
    \lim_{n \to \infty}\inf_{f \in \mathcal{F}} n^{1/2}\theta^{-}_{P^+_n,\delta}(f)  \geq 0,\\
    &\lim_{n \to \infty}\sup_{f \in \mathcal{F}} n^{1/2}\theta^{+}_{P^-_n,\delta}(f) \leq 0, \quad \quad \lim_{n \to \infty}\inf_{f \in \mathcal{F}} n^{1/2}\theta^{-}_{P^-_n,\delta}(f) < 0.
\end{align*}
Intuitively, we can for example view $P^+_n$ as a sequence of probability distributions compatible with scenario 1 above as $\sup_{f \in \mathcal{F}}\theta^+_{P^+_n,\delta}(f)$ approaches zero from above at an $n^{-1/2}$-rate. In principle $\inf_{f \in \mathcal{F}}\theta^-_{P^+_n,\delta}(f)$ can approach zero from below, so long as it is at a rate faster than $n^{-1/2}$ (our test would be unable to distinguish this from the null).

We will show that our proposed test achieves asymptotic type I error control under sampling from either $P^+_n$ or $P^-_n$. This requires us to first establish a generic weak convergence of our estimators $\{n^{1/2}\theta^+_{n,\delta}(f):f\in \mathcal{F}\}$ and $\{n^{1/2}\theta^-_{n,\delta}(f):f\in \mathcal{F}\}$ under such a sequence.
The following theorem provides conditions under which the desired weak convergence holds.

\begin{theorem}(Weak convergence under local data generating laws: qualitative heterogeneity)\label{theorem:weak_qual}
Revisiting the set-up of Lemma \ref{lemmma:local}, suppose that data are drawn as an i.i.d. triangular array $Z_{n,1},...,Z_{n,n}$ from a sequence $P_n$ that satisfies \eqref{eq:local_alt_general}. Then under the conditions of Theorem \ref{theorem:al}, 
\begin{align*}
    \theta^+_{n,\delta}(f) = \frac{1}{n}\sum_{i=1}^n \varphi^+_{0,\delta}(Z_{n,i};f) + r^+_n(f),
    \\
    \theta^-_{n,\delta}(f) = \frac{1}{n}\sum_{i=1}^n \varphi^-_{0,\delta}(Z_{n,i};f) + r^-_n(f),
\end{align*}
where $\sup_{f \in \mathcal{F}}|n^{1/2}r_n^+(f)|$ and $\sup_{f \in \mathcal{F}}|n^{1/2}r_n^-(f)|$ converge to zero in probability under sampling from $P_n$. Moreover,
$n^{1/2}\theta^+_{n,\delta}(f)$ and $n^{1/2}\theta^-_{n,\delta}(f)$, respectively, converge weakly under $P_n$ to $ \{\mathbb{G}^+(f) + c^+(f)\}$ and $\{\mathbb{G}^-(f) + c^-(f)\}$ as elements of $\ell^\infty(\mathcal{F})$, where
$\mathbb{G}^+$ and $\mathbb{G}^-$ are tight, mean zero Gaussian processes, with covariances 
\begin{align*}
    &\text{Cov}(\mathbb{G}^+(f_1), \mathbb{G}^+(f_2)) = E\{\varphi^+_{\delta}(f_1)\varphi^+_{\delta}(f_2)\},
    \\
    &\text{Cov}(\mathbb{G}^-(f_1), \mathbb{G}^-(f_2)) = E\{\varphi^-_{\delta}(f_1)\varphi^-_{\delta}(f_2)\}.
    \end{align*}
\end{theorem}

We are now able to establish type I error control under $P^+_n$ and $P^-_n$, justifying the use of our proposed test in small-sample small-signal settings.
This claim is made formal in the following lemma.
\begin{theorem}
(Type I error control under local data-generating laws: qualitative heterogeneity)\label{theorem:qual_t1}
    Assume the setting of Lemma \ref{lemmma:local} and Theorem \ref{theorem:weak_qual}, and let $t^+_{\alpha}$ and $t^-_{\alpha}$, respectively, be the $1-\alpha$  and $\alpha$  quantiles of $\sup_{f \in \mathcal{F}} \mathbb{G}(f)$ and $\inf_{f \in \mathcal{F}} \mathbb{G}(f)$. 
    Then under sampling from $P^+_n$,
    \begin{align*}
        &\limsup_{n \to \infty} P^+_n\left(n^{1/2}\sup_{f \in \mathcal{F}} \theta^+_{n,\delta}(f) >  t^+_\alpha \text{ and } n^{1/2}\inf_{f \in \mathcal{F}} \theta^-_{n,\delta}(f) < t^-_\alpha\right) \leq \alpha.
    \end{align*}
    Similarly, under sampling from $P^-_n$,
    \begin{align*}
        &\limsup_{n \to \infty} P^-_n\left(n^{1/2}\sup_{f \in \mathcal{F}} \theta^+_{n,\delta}(f) > t^+_\alpha \text{ and } n^{1/2}\inf_{f \in \mathcal{F}} \theta^-_{n,\delta}(f) <  t^-_\alpha\right) \leq \alpha.
    \end{align*}
\end{theorem}
Hence type I error is upper bounded by $\alpha$ and so for certain data-generating processes may be below $\alpha$, indicating conservative behaviour at these laws. 

\subsubsection{Power against local alternatives}

We now define the score function $\tilde{S}:\mathcal{Z} \to \mathbb{R}$ under $P_0$ that satisfies:
\begin{align}
    \sup \int \tilde{S}(z) \varphi^+_{0,\delta}(z;f)dP_0(z) > 0,  \quad 
    \inf \int \tilde{S}(z) \varphi^-_{0,\delta}(z;f)dP_0(z) < 0. \label{eq:score_tilde}
\end{align}
Then $\tilde{P}_n$ is a sequence of probability distributions that approaches $P_0$ from the path $\tilde{S}$, such that
\begin{align*}
&\lim_{n\to \infty}\int \left[ n^{1/2}\{d\tilde{P}_n(z)^{1/2}-dP_0(z)^{1/2}\}-\frac{1}{2}\tilde{S}(z)dP_0(z)^{1/2}\right]^2 = 0. 
\end{align*}
It follows furthermore from Lemma \ref{lemmma:local} that this essentially implies that 
\begin{align*}
&\lim_{n \to \infty}\sup_{f\in \mathcal{F}} \theta^{+}_{\tilde{P}_n,\delta}(f)=0, \quad \quad \lim_{n \to \infty}\inf_{f \in \mathcal{F}}  \theta^{-}_{\tilde{P}_n,\delta}(f)=0\end{align*}
and 
\begin{align*}
    &\lim_{n \to \infty}\sup_{f\in \mathcal{F}} n^{1/2}\theta^{+}_{\tilde{P}_n,\delta}(f) > 0, \quad \quad 
    \lim_{n \to \infty}\inf_{f \in \mathcal{F}} n^{1/2}\theta^{-}_{\tilde{P}_n,\delta}(f) < 0.
\end{align*}
Hence, we are considering sequences of distributions such that alternative holds at any finite $n$, but becomes more challenging to detect (in the sense of shrinking closer to the null) as sample size increases. 

The following result is a further consequence of Theorems \ref{theorem:weak_qual}: 
\begin{theorem}
(Power against local alternatives: qualitative heterogeneity)\label{theorem:power_qual_local}
    Assume the setting of Lemma \ref{lemmma:local} and Theorem \ref{theorem:weak_qual}, and let $t^+_{\alpha}$ and $t^-_{\alpha}$, respectively, be the $(1-\alpha)$ and $\alpha$  quantiles of $\sup_{f \in \mathcal{F}} \mathbb{G}^+(f)$ and $\inf_{f \in \mathcal{F}} \mathbb{G}^-(f)$.
    Then under sampling from sequences $\tilde{P}_n$ that satisfy \eqref{eq:score_tilde},
    \begin{align*}
        &\liminf_{n \to \infty} \tilde{P}_n\left(n^{1/2}\sup_{f \in \mathcal{F}} \theta^+_{n,\delta}(f) >  t^+_\alpha \text{ and } n^{1/2}\inf_{f \in \mathcal{F}} \theta^-_{n,\delta}(f) < t^-_\alpha\right)\\
        & \geq \max\left\{0, P_0\left(\sup_{f\in \mathcal{F}}\{\mathbb{G}^+(f) + c^+(f)\} > t_\alpha^{+}\right)+P_0\left(\inf_{f \in \mathcal{F}}\{\mathbb{G}^-(f) + c^-(f)\} < t_\alpha^{-}\right)-1\right\}.
    \end{align*}
\end{theorem}
We are therefore unable to guarantee non-trivial power in general against certain classes of local alternatives. When $\sup_{f \in \mathcal{F}} \theta^+_{0,\delta}(f)$ and $\inf_{f \in \mathcal{F}}\theta^-_{0,\delta}(f)$ are close enough to zero (relative to the sample size), then it may be that power is lower than $\alpha$. This is unsurprising, given that even in a simple setting where $X$ is a binary variable and subgroup effects are uncorrelated, the likelihood ratio test is know to suffer from low power in such cases. This is in spite of it being optimal within the class of monotone tests \citep{berger1989uniformly}. In this particular setting, if one chooses $\mathcal{F}$ as the set of indicator functions indicating membership of a subgroup, our test approximately coincides with test of \citet{piantadosi1993comparison}, which is essentially equivalent to the likelihood ratio test. The approximation occurs due to our use of the multiplier bootstrap for characterizing the distribution of the test statistic. Hence we would expect that our power at least is acceptable relative to alternatives in simple settings. 

On the other hand, if $c^+(f)=+\infty$ or $c^{-}(f)=-\infty$ for some $f$, then power is asymptotically bounded away from $\alpha$. Hence the power of our test exceeds $\alpha$ when either of the one-sided tests is rejected with probability nearly equal to one. For instance, suppose the CATE greatly exceeds $\delta=0$ for some subgroups and is in a neighborhood of zero for other subgroups (i.e., $\theta^+$ is large while $\theta^-$ is nearly zero).
In this case, we would expect to reject $\theta^+ \leq 0$ with probability nearly equal to one, so the power to detect qualitative heterogeneity will be determined by our power to reject the null that $\theta^- \geq 0$, which is larger than $\alpha$. We have focused on local asymptotics for the sub-null that $\sup_{f \in \mathcal{F}}\theta^{+}_{0,\delta}(f) = \inf_{f\in\mathcal{F}}\theta^{-}_{0,\delta}(f) = 0$ since this is the most-challenging setting.

Unlike \citet{shi2019sparse}, our results on local asymptotic power do not appear to require a `margin condition' \citep{luedtke2016statistical}, which restricts the amount of mass that $\tau(X_s)$ is allowed to have at zero. On the other hand, they are restricted to $\mathcal{F}$ being a Donsker class. It is an interesting question whether a margin condition or other similar restrictions could lead to improvements in power when testing qualitative heterogeneity.

\section{Proofs of main results}\label{appendix:proofs}

\subsection{Proof of Lemma \ref{lemma:eif}}

\begin{proof}
We first give the result for $\theta^{+}_{P,\tau_P}(f)$
(the one for $\theta^{-}_{P,\tau_P}(f)$ follows along the same lines). Let $P_t$ be a parametric submodel indexed by parameter $t$, with associated density $p_t(Z)$. As in the main text, to simplify notation we will index quantities that depend on $P_t$ by $t$ rather than $P_t$. Then the score function is defined as 
\[\frac{\partial}{\partial t}\log p_t(Z)|_{t=0}=S(Z).\]
Our goal is to find $\varphi^+_{P,\tau_P}(Z;f)$, where 
\[\frac{\partial}{\partial t}\theta^{+}_{t,\tau_t}(f)|_{t=0}=E_{P}\{\varphi^+_{P,\tau_P}(Z;f)S(Z)\}.\] 

First,
\begin{align*}
\frac{\partial }{\partial t}\theta^{+}_{t,\tau_t}(f)|_{t=0}&=\frac{\partial}{\partial t} E_t[\{\mu_t(1,X)-\mu_t(0,X)-\tau_{t}\}f(X_s)]|_{t=0}\\
&=\frac{\partial}{\partial t} E_t[\{\mu_P(1,X)-\mu_P(0,X)-\tau_P\}f(X_s)]|_{t=0}\\
&\quad+ E_{P}[\partial \{\mu_t(1,X)-\mu_t(0,X)\} /\partial t|_{t=0}f(X_s)]- \partial \tau_{t}/\partial t|_{t=0}E_{P}\{f(X_s)\}.
\end{align*}
It can be shown that 
\begin{align*}
&\frac{\partial}{\partial t} E_t[\{\mu_P(1,X)-\mu_P(0,X)-\tau_{P}\}f(X_s)]|_{t=0}\\
&=E_P\left([\{\mu_P(1,X)-\mu_P(0,X)-\tau_P\}f(X_s)-\theta^{+}_{P,\tau_P}(f)]S(Z)\right).
\end{align*}
Further,
\begin{align*}
E_{P}[\partial \{\mu_t(1,X)-\mu_t(0,X)\} /\partial t|_{t=0}f(X_s)]&=E_P \left[\frac{(2A-1)}{\pi_P(A|X)}\{Y-\mu_P(A,X)\}f(X_s)S(Z)\right]
\end{align*}
and 
\begin{align*}
&-\partial \tau_{t}/\partial t|_{t=0}E_{P}\{f(X_s)\}\\
&=-E_P \left(\left[\frac{(2A-1)}{\pi_P(A|X)}\{Y-\mu_P(A,X)\} +\mu_P(1,X)-\mu_P(0,X)-\tau_P\right]S(Z)\right)E_{P}\{f(X_s)\}.
\end{align*}
Combining these terms gives us the influence function:
\begin{align*}
\left[\frac{(2A-1)}{\pi_P(A|X)}\{Y-\mu_P(A,X)\} +\mu_P(1,X)-\mu_P(0,X)-\tau_P\right]\left[f(X_s)-E_P\{f(X_s)\}\right]-\theta^{+}_{P,\tau_P}.
\end{align*}
To obtain the result for $\theta^{+}_{P,\delta}(f)$ and $\theta^{-}_{P,\delta}(f)$, one can repeat the previous arguments, replacing $\tau_P$ with $\delta$ which is now fixed.
\end{proof}

\subsection{Proof of Theorem \ref{theorem:al}}

\begin{proof}
We will show the result for $\theta^+_{n,\tau_n}(f)$. For a fixed $f$, we have that 
\begin{align*}
     r^+_{n,\tau_n}(f)=R_1(f)+R_2(f)
\end{align*}
where 
\begin{align*}
R_1(f)&:=  \frac{1}{n}\sum_{i=1}^n\left[\left\{\psi_n(Z_i)-\tau_n\right\}\{f(X_{s,i}) - \bar{f}_n\}-\left\{\psi_0(Z_i)-\tau_0\right\}\left\{f(X_{s,i})-\bar{f}_0\right\}\right]\\&
-\int \left[\left\{\psi_n(z)- \tau_n\right\}\{f(x_{s})-\bar{f}_n\}-\left\{\psi_0(z)-\tau_0\right\}\left\{f(x_{s})-\bar{f}_0\right\}\right]dP_0(z),\\
R_2(f)&:=\int\left[ \left\{\psi_n(z) - \tau_n \right\}\{f(x_{s})-\bar{f}_n\} - \theta^+_{0,\tau_0}(f)\right]dP_0(z),
\end{align*}
$\bar{f}_n=n^{-1}\sum_{i=1}^n f(X_{s,i})$ and $\bar{f}_0=E_0\{f(X_{s})\}$. 

Considering first $R_1(f)$, note that 
it follows from Assumptions \ref{donsker}, \ref{consist} and the regularity conditions of the theorem that 
\[\int \left[\left\{\psi_n(z)- \tau_n\right\}\{f(x_{s})-\bar{f}_n\}-\left\{\psi_0(z)-\tau_0\right\}\left\{f(x_{s})-\bar{f}_0\right\}\right]^2dP_0(z)=o_{P_0}(1).\]
As a consequence of the Donsker class condition in Assumption \ref{donsker}, this also holds uniformly over $\mathcal{F}$:
\[\sup_{f\in\mathcal{F}}\int \left[\left\{\psi_n(z)- \tau_n\right\}\{f(x_{s})-\bar{f}_n\}-\left\{\psi_0(z)-\tau_0\right\}\left\{f(x_{s})-\bar{f}_0\right\}\right]^2dP_0(z)=o_{P_0}(1).\]
By invoking Assumption \ref{donsker} again, this result implies that \[\sup_{f\in \mathcal{F}}|R_1(f)|=o_{P_0}(n^{-1/2})\] by Lemma 19.24 and arguments from the proof of Theorem 19.26 in \citet{van2000asymptotic}.

Moving onto $R_{2}(f)$, then
\begin{align*}
R_2(f)&=\int\left[ \left\{\psi_n(z) - \tau_n \right\}\{f(x_{s})-\bar{f}_n\} - \{\mu_0(1,x)-\mu_0(0,x)-\tau_0\}\{f(x_s)-\bar{f}_0\}\right]dP_0(z)\\
&=\int \left[\psi_n(z)-\{\mu_0(1,x)-\mu_0(0,x)\}\right]f(x_{s})dP_0(z)\\
&\quad - (\tau_n-\tau_0)\bar{f}_0-\left[\int  \{\psi_n(z)-\tau_n\}dP_0(z)\right]\bar{f}_n+\left[\int \{\mu_0(1,x)-\mu_0(0,x)-\tau_0\}dP_0(z)\right]\bar{f}_0\\
&=\int \left[\psi_n(z)-\{\mu_0(1,x)-\mu_0(0,x)\}\right]f(x_{s})dP_0(z)\\
&\quad - (\tau_n-\tau_0)\bar{f}_0-\left[\int  \{\psi_n(z)-\tau_n\}dP_0(z)\right]\bar{f}_0-\left[\int  \{\psi_n(z)-\tau_n\}dP_0(z)\right](\bar{f}_n-\bar{f}_0)\\
&=\int \left[\psi_n(z)-\{\mu_0(1,x)-\mu_0(0,x)\}\right]f(x_{s})dP_0(z)\\
&\quad -\left[\int \{\psi_n(z)-\tau_0\}dP_0(z)\right]\bar{f}_0-\left[\int  \{\psi_n(z)-\tau_n\}dP_0(z)\right](\bar{f}_n-\bar{f}_0)\\
&=\int \left[\psi_n(z)-\{\mu_0(1,x)-\mu_0(0,x)\}\right]f(x_{s})dP_0(z)\\
&\quad -\left[\int \{\psi_n(z)-\tau_0\}dP_0(z)\right]\bar{f}_0-\left[\int  \{\psi_n(z)-\tau_n\}dP_0(z)\right](\bar{f}_n-\bar{f}_0)\\&\quad +\left[\int \{\mu_0(1,x)-\mu_0(0,x)-\tau_0\}dP_0(x)\right](\bar{f}_n-\bar{f}_0)\\
&=\int \left[\psi_n(z)-\{\mu_0(1,x)-\mu_0(0,x)\}\right]f(x_{s})dP_0(z)\\
&\quad -\left(\int \left[\psi_n(z)-\{\mu_0(1,x)-\mu_0(0,x)\}\right]dP_0(z)\right)\bar{f}_0-\left[\int  \{\psi_n(z)-\tau_n\}dP_0(z)\right](\bar{f}_n-\bar{f}_0)\\&\quad +\left[\int \{\mu_0(1,x)-\mu_0(0,x)-\tau_0\}dP_0(z)\right](\bar{f}_n-\bar{f}_0)\\
&=\int \left[\psi_n(z)-\{\mu_0(1,x)-\mu_0(0,x)\}\right]\{f(x_{s})-\bar{f}_0\}dP_0(z)\\
& \quad + \left(\int \left[\psi_n(z)-\{\mu_0(1,x)-\mu_0(0,x)\}\right]dP_0(z)\right)(\bar{f}_0-\bar{f}_n)\\
& \quad + (\tau_n-\tau_0)(\bar{f}_n-\bar{f}_0)\\
&=R_{2_{(i)}}(f)+R_{2_{(ii)}}(f)+R_{2_{(iii)}}(f)
\end{align*}
where the second equality follows from the definition of $\bar{f}_0$, the third and the fifth because $\int \{\mu_0(1,x)-\mu_0(0,x)-\tau_0\}dP_0(x)=0$ and the fourth  through cancellation and rearrangement of terms. We will now show that each of the terms in the final expression converge uniformly over $\mathcal{F}$ to zero at a rate $n^{-1/2}$.

First,
\begin{align*}
R_{2_{(i)}}(f)&=\sum^1_{a=0}\int (-1)^{1+a}\left\{\pi_0(a|x)-\pi_n(a|x) \right\}\left\{\mu_0(a,x)-\mu_n(a,x)\right\}\pi^{-1}_n(a|x) \{f(x_{s})-\bar{f}_0\}dP_0(x).
\end{align*}
By Assumption \ref{product}, the assumption that $\pi_n(a|X)$ is bounded below with probability 1 and that $f(X_s)$ lies in $[0,1]$, it follows by application of the Cauchy-Schwarz inequality that 
\begin{align*}
\bigg|\sum^1_{a=0}\int (-1)^{1+a}\left\{\pi_0(a|x)-\pi_n(a|x) \right\}\left\{\mu_0(a,x)-\mu_n(a,x)\right\}\pi^{-1}_n(a|x) \{f(x_{s})-\bar{f}_0\}dP_0(x)\bigg|=o_{P_0}(n^{-1/2})
\end{align*}
and therefore $R_{2_{(i)}}(f)=o_{P_0}(n^{-1/2})$. 

For $R_{2_{(ii)}}(f)$, similar arguments establish that 
\[\bigg|\int \left[\psi_n(z)-\{\mu_0(1,x)-\mu_0(0,x)\}\right]dP_0(z)\bigg|=o_{P_0}(n^{-1/2}).\]
It is straightforward that $\bar{f}_n-\bar{f}=O_{P_0}(n^{-1/2})$ since $f$ is fixed. By Slutsky's Theorem, it follows that $R_{2_{(ii)}}(f)=o_{P_0}(n^{-1/2})$. Moreover, one can establish under the same conditions that 
\[R_{2_{(iii)}}(f)=o_{P_0}(1)O_{P_0}(n^{-1/2})=o_{P_0}(n^{-1/2}).\] 

Finally, following the proof of Theorem 19.26 in \citet{van2000asymptotic}, it follows by Assumption \ref{donsker} that $\sup_{f\in \mathcal{F}}|R_{2_{(i)}}(f)|=o_{P_0}(n^{-1/2})$, $\sup_{f\in \mathcal{F}}|R_{2_{(ii)}}(f)|=o_{P_0}(n^{-1/2})$ and $\sup_{f\in \mathcal{F}}|R_{2_{(iii)}}(f)|=o_{P_0}(n^{-1/2})$. Then by repeated application of the triangle inequality, we have that $\sup_{f\in \mathcal{F}}|R_{2}(f)|=o_{P_0}(n^{-1/2})$ and moreover that $\sup_{f\in \mathcal{F}}|r^+_{n,\tau_n}(f)|=o_{P_0}(n^{-1/2})$.

We note that an equivalent result can be shown for $\theta^-_{n,\tau_n}(f)$. For $\theta^+_{0,\delta}(f)$ (and $\theta^-_{0,\delta}(f)$) the result could be established under a simplification of the proceeding proof, which is omitted for brevity.
\end{proof}

\subsection{Proof of Theorem \ref{theorem:size-quant}}

\begin{proof}
Under the null hypothesis, $\theta^+_{0,\tau_0}(f)-\theta^-_{0,\tau_0}(f)=0$ $\forall f\in \mathcal{F}$. Then following Theorem \ref{theorem:al} and Corollary \ref{corollary:weak}, $n^{1/2}\{\theta^+_{n,\tau_n}(f)-\theta^-_{n,\tau_n}(f)\}$ converges in distribution to a mean-zero Gaussian random variable, pointwise in $f$. Moreover, $n^{1/2}\{\theta^+_{n,\tau_n}(f)-\theta^-_{n,\tau_n}(f)\}$ converges weakly in $\ell^\infty(\mathcal{F})$ to $\mathbb{G}(f)$. By application of the continuous mapping theorem, $n^{1/2}\sup_{f\in \mathcal{F}}|\theta^+_{n,\tau_n}(f)-\theta^-_{n,\tau_n}(f)|$ converges in distribution to $\sup_{f\in \mathcal{F}}|\mathbb{G}(f)|$; here we use the uniform continuity of the supremum map on $\ell^\infty(\mathcal{F})$. Hence 
\begin{align*}
&\lim_{n\to \infty} P_0\left(n^{1/2}\sup_{f\in \mathcal{F}} |\theta^+_{n,{\tau_n}}(f)-\theta^-_{n,{\tau_n}}(f)|> t_\alpha \right)\\
&=P_0\left(\sup_{f\in \mathcal{F}}|\mathbb{G}(f)|> t_\alpha \right)\\&=\alpha.
\end{align*}
\end{proof}

\subsection{Proof of Theorem \ref{theorem:size-qual}}

\begin{proof}
Suppose first that $\sup_{f\in \mathcal{F}} \theta^+_{0,\delta}(f) >0$ and $\inf_{f\in \mathcal{F}}\theta^-_{0,\delta}(f)\geq 0$. By Corollary \ref{corollary:weak} and the continuous mapping theorem, $n^{1/2}\inf_{f\in \mathcal{F}}\{\theta^-_{n,\delta}(f)-\theta^-_{0,\delta}(f)\}$ converges in distribution to $\inf_{f\in \mathcal{F}}\mathbb{G}^-(f)$. Also, 
$n^{1/2}\sup_{f\in \mathcal{F}}\{\theta^+_{n,\delta}(f)-\theta^+_{0,\delta}(f)\}$ converges in distribution to $\sup_{f\in \mathcal{F}}\mathbb{G}^+(f)$. We have that
\begin{align*}
P_0\left(n^{1/2} \sup_{f \in \mathcal{F}} \theta^+_{n,\delta}(f) > t^+_\alpha\right)
&\geq P_0\left(n^{1/2} \sup_{f \in \mathcal{F}}\{\theta^+_{n,\delta}(f)-\theta^+_{0,\delta}(f)\} > t^+_\alpha-n^{1/2} \sup_{f \in \mathcal{F}} \theta^+_{0,\delta}(f)\right)
\end{align*}
using the sub-additivity of suprema; the right hand side tends to one by Slutsky's theorem, since $n^{1/2} \sup_{f \in \mathcal{F}} \theta^+_{0,\delta}(f)$ diverges to $+\infty$. Also,
\begin{align*}
P_0\left(n^{1/2} \inf_{f \in \mathcal{F}} \theta^-_{n,\delta}(f) < t^-_\alpha\right)&\leq  P_0\left(n^{1/2} \inf_{f \in \mathcal{F}}\theta^-_{n,\delta}(f)-n^{1/2}\inf_{f \in \mathcal{F}}\theta^-_{0,\delta}(f) < t^-_\alpha\right)\\
&\leq  P_0\left(n^{1/2} \inf_{f \in \mathcal{F}}\{\theta^-_{n,\delta}(f)-\theta^-_{0,\delta}(f)\} < t^-_\alpha\right)
\end{align*}
where we use that $\inf_{f \in \mathcal{F}} \theta^-_{0,\delta}(f)\geq 0$ and the sub-additivity of infima. Then
\begin{align*}
\lim_{n\to\infty} P_0\left(n^{1/2} \inf_{f \in \mathcal{F}}\{\theta^-_{n,\delta}(f)-\theta^-_{0,\delta}(f)\} < t^-_\alpha\right)&=P_0\left(\inf_{f \in \mathcal{F}}\mathbb{G}^-(f) < t^-_\alpha\right)\\
&=\alpha.
\end{align*}
As a consequence, 
\begin{align*}
\lim_{n \to \infty} P_0\left(n^{1/2} \sup_{f \in \mathcal{F}} \theta^+_{n,\delta}(f) > t^+_\alpha\right)&=1,\\
\limsup_{n \to \infty} P_0\left(n^{1/2} \inf_{f \in \mathcal{F}} \theta^-_{n,\delta}(f) < t^-_\alpha\right)
&\leq \alpha.
\end{align*}
Furthermore, by the sub-additivity of the limit superior,
\begin{align}\label{eq:lim_min}
\limsup_{n\to \infty}\min\left\{P_0\left(n^{1/2} \sup_{f \in \mathcal{F}} \theta^+_{n,\delta}(f) > t^+_\alpha\right), P_0\left(n^{1/2} \inf_{f \in \mathcal{F}}  \theta^-_{n,\delta}(f) < t^-_\alpha\right)\right\}\leq\min(\alpha,1)=\alpha.
\end{align}

Using the Fr\'echet inequalities, the probability of rejecting the null of no qualitative heterogeneity can be upper bounded by:
\begin{align*}
    &P_0\left(n^{1/2}\sup_{f \in \mathcal{F}}  \theta^+_{n,\delta}(f) > t^+_\alpha \text{ and } n^{1/2}\inf_{f \in \mathcal{F}}\theta^-_{n,\delta}(f) < t^-_\alpha\right)\\
    &\leq \min\left\{P_0\left(n^{1/2} \sup_{f \in \mathcal{F}} \theta^+_{n,\delta}(f) > t^+_\alpha\right), P_0\left(n^{1/2} \inf_{f \in \mathcal{F}}  \theta^-_{n,\delta}(f) < t^-_\alpha\right)\right\}.
\end{align*}
Since this holds for all $n$, then by \eqref{eq:lim_min},
\begin{align*}
    &\limsup_{n\to \infty} P_0\left(n^{1/2}\sup_{f \in \mathcal{F}} \theta^+_{n,\delta}(f) > t^+_\alpha \text{ and } n^{1/2}\inf_{f \in \mathcal{F}}  \theta^-_{n,\delta}(f) < t^-_\alpha\right)\\
    &\leq \limsup_{n\to \infty}\min\left\{P_0\left(n^{1/2} \sup_{f \in \mathcal{F}}\theta^+_{n,\delta}(f) > t^+_\alpha\right), P_0\left(n^{1/2} \inf_{f \in \mathcal{F}} \theta^-_{n,\delta}(f) < t^-_\alpha\right)\right\}\\
     &\leq\alpha.
     \end{align*}

Using essentially identical arguments, if $\sup_{f\in \mathcal{F}} \theta^+_{0,\delta}(f)\leq 0$ while $\inf_{f\in \mathcal{F}}\theta^-_{0,\delta}(f)<0$, then asymptotically the rejection rate is again upper bounded by $\min(\alpha,1)=\alpha$. Moreover, if $\sup_{f\in \mathcal{F}} \theta^+_{0,\delta}(f)=\inf_{f\in \mathcal{F}}\theta^-_{0,\delta}(f)=0$, then the asymptotic rejection rate is upper bounded by $\min(\alpha,\alpha)=\alpha$. Since these three possibilities exhaust the null, then the main result follows.

\end{proof}

\subsection{Proof of Theorem \ref{theorem:power_f-quant}}

\begin{proof}

By Theorem \ref{theorem:al}, Corollary \ref{corollary:weak} and an application of the reverse triangle inequality, one can also establish that $\sup_{f\in \mathcal{F}}|\theta^+_{n,\tau_n}(f)-\theta^-_{n,\tau_n}(f)|$ converges in probability to $\sup_{f\in \mathcal{F}}|\theta^+_{0,\tau_0}(f)-\theta^-_{0,\tau_0}(f)|$. Now since there exists at least one $f\in\mathcal{F}$ where $\theta^+_{0,\tau_0}(f)-\theta^-_{0,\tau_0}(f)\neq 0$, then $\sup_{f\in \mathcal{F}}|\theta^+_{0,\tau_0}(f)-\theta^-_{0,\tau_0}(f)|>0$. By the consistency result, it follows that $n^{1/2}\sup_{f\in \mathcal{F}}|\theta^+_{n,\tau_n}(f)-\theta^-_{n,\tau_n}(f)|$ diverges to positive infinity. The result follows after taking the limit of the rejection rate.

\end{proof}

\subsection{Proof of Theorem \ref{theorem:power_f-qual}}

\begin{proof}
Following the arguments in the proof of Theorem \ref{theorem:size-quant}, if $\sup_{f \in \mathcal{F}} \theta^+_{0,\delta}(f) > 0$ and $\inf_{f \in \mathcal{F}} \theta^-_{0, \delta}(f) < 0$, a consequence of Theorem \ref{theorem:al} and Corollary \ref{corollary:weak} is that 
\begin{align*}
&\lim_{n \to \infty} P_0\left(n^{1/2} \sup_{f \in \mathcal{F}}  \theta^+_{n,\delta}(f) > t^+_\alpha\right)=1,\\
&\lim_{n \to \infty} P_0\left(n^{1/2} \inf_{f \in \mathcal{F}}  \theta^-_{n,\delta}(f) < t^-_\alpha\right)=1.
\end{align*}
Then
\begin{align*}
&\lim_{n\to \infty} \left\{P_0\left(n^{1/2}\sup_{f \in \mathcal{F}}\theta^+_{n,\delta}(f) > t^+_\alpha\right)+P_0\left(n^{1/2} \inf_{f \in \mathcal{F}} \theta^-_{n,\delta}(f)< t^-_\alpha\right)-1\right\} \\
&=\lim_{n\to \infty} P_0\left(n^{1/2}\sup_{f \in \mathcal{F}}  \theta^+_{n,\delta}(f) > t^+_\alpha\right)+\lim_{n\to \infty} P_0\left(n^{1/2} \inf_{f \in \mathcal{F}} \theta^-_{n,\delta}(f)< t^-_\alpha\right)-1\\
&=1.
\end{align*}
Furthermore, 
\begin{align}\label{eq:lim_max}
\lim_{n\to \infty} \max\bigg\{0, P_0\left(n^{1/2}\sup_{f \in \mathcal{F}}  \theta^+_{n,\delta}(f) > t^+_\alpha\right)+P_0\left(n^{1/2} \inf_{f \in \mathcal{F}} \theta^-_{n,\delta}(f)< t^-_\alpha\right)-1
    \bigg\}=1.
\end{align}

Now by the Fr\'echet inequalities, for any $n$ we have
\begin{align*}
    &P_0\left(n^{1/2}\sup_{f \in \mathcal{F}}  \theta^+_{n,\delta}(f) > t^+_\alpha \text{ and } n^{1/2} \inf_{f \in \mathcal{F}} \theta^-_{n,\delta}(f) < t^-_\alpha\right)
    \\
 &\geq \max\bigg\{0, P_0\left(n^{1/2}\sup_{f \in \mathcal{F}}  \theta^+_{n,\delta}(f) > t^+_\alpha\right)+P_0\left(n^{1/2} \inf_{f \in \mathcal{F}} \theta^-_{n,\delta}(f)< t^-_\alpha\right)-1
    \bigg\}.
\end{align*}
The result then follows by applying the squeeze theorem in combination with \eqref{eq:lim_max}.

\end{proof}

\subsection{Proof of Theorem \ref{theorem:local_quant}}

\begin{proof}
Under $P_0$, we have that $\theta^+_{0,\tau_0}(f)-\theta^-_{0,\tau_0}(f)=0$ $\forall f \in \mathcal{F}$. A direct consequence of Theorem \ref{theorem:al} is that 
\[\sup_{f\in\mathcal{F}}\bigg|n^{1/2}\{\theta^+_{n,\tau_n}(f)-\theta^-_{n,\tau_n}(f)\}-n^{-1/2}\sum_{i=1}^n \{\varphi^+_{0,\tau_0}(Z_{n,i};f)-\varphi^-_{0,\tau_0}(Z_{n,i};f)\}\bigg|\overset{P_0}{\rightarrow}0.\]
  It follows from Lemma 3.10.11 of \citet{vanderVaart1996} that $P_n$ is contiguous with respect to $P_0$ under \eqref{eq:local_alt}. Hence by Theorem 3.10.5 of \citet{vanderVaart1996}, we have that
  \[\sup_{f\in\mathcal{F}}\bigg|n^{1/2}\{\theta^+_{n,\tau_n}(f)-\theta^-_{n,\tau_n}(f)\}-n^{-1/2}\sum_{i=1}^n \{\varphi^+_{0,\tau_0}(Z_{n,i};f)-\varphi^-_{0,\tau_0}(Z_{n,i};f)\}\bigg|\overset{P_n}{\rightarrow}0.\]
  Finally, under the Donsker condition in Assumption \ref{donsker}, Theorem 3.10.12 of \citet{vanderVaart1996} implies that 
  \[\left\{n^{-1/2}\sum_{i=1}^n \{\varphi^+_{0,\tau_0}(Z_{n,i};f)-\varphi^-_{0,\tau_0}(Z_{n,i};f)\}:f\in \mathcal{F}\right\}\]
  converges to $\{\mathbb{G}(f)+c(f):f\in\mathcal{F}\}$ as an element in $\ell^\infty(\mathcal{F})$.
  
\end{proof}

\subsection{Proof of Corollary \ref{corollary:local_quant}}

\begin{proof}
By Theorem \ref{theorem:local_quant} and the continuous mapping theorem, we have that $n^{1/2}\sup_{f\in \mathcal{F}}|\theta^+_{n,\tau_n}(f)-\theta^-_{n,\tau_n}(f)|$ converges in distribution to $\sup_{f\in\mathcal{F}}|\mathbb{G}(f)+c(f)|$ under $P_n$. Therefore 
\begin{align*}
 &\lim_{n \to \infty} P_n \left(n^{1/2}\sup_{f\in \mathcal{F}}|\theta^+_{n,\tau_n}(f)-\theta^-_{n,\tau_n}(f)|>t_\alpha\right)\\
 &=P_0 \left(\sup_{f\in\mathcal{F}}|\mathbb{G}(f)+c(f)|>t_\alpha\right)\\&>\alpha.
\end{align*}
\end{proof}

\subsection{Proof of Lemma \ref{lemmma:local}}

\begin{proof}
We first observe that because $P_n$ approaches $P_0$, we have $\sup_{f \in \mathcal{F}} \theta^+_{P_n,\delta}(f)$ and $\inf_{f \in \mathcal{F}} \theta^-_{P_n,\delta}(f)$ both tend to zero in the limit of large $n$, since $\sup_{f \in \mathcal{F}} \theta^+_{0,\delta}(f)=\inf_{f \in \mathcal{F}} \theta^-_{0,\delta}(f)=0$.

Furthermore, 
\begin{align*}
\theta^+_{P_n,\delta}(f)=&\int \varphi_{0,\delta}^+(z;f)\{dP_n(z)-dP_0(z)\}+R^+(P_n,P_0),\\
n^{1/2}\theta^+_{P_n,\delta}(f)=&\int S(z)\varphi_{0,\delta}^+(z;f)dP_0(z)+\int \varphi_{0,\delta}^+(z;f)\{n^{1/2}dP_n(z)-n^{1/2}dP_0(z)-S(z)dP_0(z)\}\\&+n^{1/2}R^+(P_n,P_0).
\end{align*}
Firstly, $n^{1/2}R^+(P_n,P_0)$ converges uniformly in $f$ to zero by assumption. Furthermore, following the proof of Theorem 3.10.12 in \citet{vanderVaart1996}, \eqref{eq:local_alt_general} implies that 
\[\int \varphi_{0,\delta}^+(z;f)\{n^{1/2}dP_n(z)-n^{1/2}dP_0(z)-S(z)dP_0(z)\}\]
also converges to zero uniformly in $f$. This implies the first part of \eqref{eq:lim_score}; the second part follows using the same reasoning.
\end{proof}

\subsection{Proof of Theorem \ref{theorem:weak_qual}}

\begin{proof}
Weak convergence results can be obtained for $\theta^+_{n,\delta}(f)$ and $\theta^-_{n,\delta}(f)$ along the lines of the proof of Theorem \ref{theorem:local_quant}, noting that $\theta^+_{0,\delta}(f)$ and $\theta^-_{0,\delta}(f)=0$ $\forall f\in \mathcal{F}$. Namely, uniform asymptotic linearity under $P_0$ of $\theta^+_{n,\delta}(f)$ and $\theta^-_{n,\delta}(f)$ follows from Theorem \ref{theorem:al}, contiguity w.r.t  $P_n$ follows from Lemma 3.10.11 of \citet{vanderVaart1996}, uniform asymptotic linearity under $P_n$ follows from Theorem 3.10.5 of \citet{vanderVaart1996} and the resulting weak convergence result follows by application of Theorem 3.10.12 of \citet{vanderVaart1996}. 

\end{proof}

\subsection{Proof of Theorem \ref{theorem:qual_t1}}

\begin{proof}
For any $n$, 
\begin{align*}
&P^+_n\left(n^{1/2}\sup_{f \in \mathcal{F}} \theta^+_{n,\delta}(f) > t_\alpha^+ \text{ and } n^{1/2}\inf_{f \in \mathcal{F}} \theta^-_{n,\delta}(f) < t^-_\alpha\right) \\
&\leq P^+_n\left(n^{1/2}\inf_{f \in \mathcal{F}} \theta^-_{n,\delta}(f) < t^-_\alpha\right).
\end{align*}
By Theorem \ref{theorem:weak_qual},
\begin{align}\label{equation:inf_local_con}
 &\lim_{n \to \infty} P^+_n\left(n^{1/2}\inf_{f \in \mathcal{F}} \theta^-_{n,\delta}(f) < t_\alpha^-\right)=P_0\left(\inf_{f \in \mathcal{F}}\left\{\mathbb{G}^-(f) + c^-(f)\right\} < t_\alpha^{-} \right).
 \end{align}

Then
\begin{align*}
&\limsup_{n\to \infty} P^+_n\left(n^{1/2}\sup_{f \in \mathcal{F}} \theta^+_{n,\delta}(f) > t_\alpha^+ \text{ and } n^{1/2}\inf_{f \in \mathcal{F}} \theta^-_{n,\delta}(f) < t^-_\alpha\right) \\
&\leq \limsup_{n\to \infty} P^+_n\left(n^{1/2}\inf_{f \in \mathcal{F}} \theta^-_{n,\delta}(f) < t^-_\alpha\right)\\
&=P_0\left(\inf_{f \in \mathcal{F}}\{\mathbb{G}^-(f) + c^-(f)\} < t_\alpha^{-} \right)\\
        &\leq P_0\left(\inf_{f \in \mathcal{F}}\mathbb{G}^-(f)+\inf_{f \in \mathcal{F}}c^-(f) < t_\alpha^{-} \right) \\
    &\leq P_0\left(\inf_{f \in \mathcal{F}}\mathbb{G}^-(f) < t_\alpha^{-} \right) \\&\leq 
    \alpha
\end{align*}
where we use \eqref{equation:inf_local_con} and the fact that $\inf_{f \in \mathcal{F}}c^-(f)$ is non-negative by the restriction on the scores. A similar argument shows that type I error control is preserved under sampling from $P^-_n$.

\end{proof}

\subsection{Proof of Theorem \ref{theorem:power_qual_local}}

\begin{proof}
Under sampling from $\tilde{P}_n$, for any $n$ we have 
\begin{align*}
&\tilde{P}_n\left(n^{1/2}\sup_{f \in \mathcal{F}} \theta^+_{n,\delta}(f) > t_\alpha^+ \text{ and } n^{1/2}\inf_{f \in \mathcal{F}} \theta^-_{n,\delta}(f) <t^-_\alpha\right) \\
&\geq \max\left\{0, \tilde{P}_n\left(n^{1/2}\sup_{f \in \mathcal{F}}\theta^+_{n,\delta}(f) > t^+_\alpha\right)+\tilde{P}_n\left(n^{1/2} \inf_{f \in \mathcal{F}} \theta^-_{n,\delta}(f)< t^-_\alpha\right)-1\right\}
\end{align*}
by the Fr\'echet inequalities. By Theorem \ref{theorem:weak_qual},
\begin{align*}
 &\lim_{n \to \infty} \tilde{P}_n\left(n^{1/2}\sup_{f \in \mathcal{F}} \theta^+_{n,\delta}(f) >  t^+_\alpha\right)=P_0\left(\sup_{f \in \mathcal{F}}\left\{\mathbb{G}^+(f) + c^+(f)\right\} > t^{+}_\alpha \right)    
\end{align*}
 and hence
\begin{align*}
&\lim_{n\to \infty} \left\{\tilde{P}_n\left(n^{1/2}\sup_{f \in \mathcal{F}}\theta^+_{n,\delta}(f) > t^+_\alpha\right)+\tilde{P}_n\left(n^{1/2} \inf_{f \in \mathcal{F}} \theta^-_{n,\delta}(f)< t^-_\alpha\right)-1\right\} \\
&=P_0\left(\sup_{f\in \mathcal{F}}\{\mathbb{G}^+(f) + c^+(f)\} > t^+_\alpha\right)+P_0\left(n^{1/2} \inf_{f \in \mathcal{F}}\{\mathbb{G}^-(f) + c^-(f)\}< t^-_\alpha\right)-1
\end{align*}
also using \eqref{equation:inf_local_con}. Therefore 
\begin{align*}
&\liminf_{n\to\infty}\tilde{P}_n\left(n^{1/2}\sup_{f \in \mathcal{F}} \theta^+_{n,\delta}(f) > t_\alpha^+ \text{ and } n^{1/2}\inf_{f \in \mathcal{F}} \theta^-_{n,\delta}(f) <t^-_\alpha\right) \\
&\geq \liminf_{n\to\infty} \max\left\{0, \tilde{P}_n\left(n^{1/2}\sup_{f \in \mathcal{F}}\theta^+_{n,\delta}(f) > t^+_\alpha\right)+\tilde{P}_n\left(n^{1/2} \inf_{f \in \mathcal{F}} \theta^-_{n,\delta}(f)< t^-_\alpha\right)-1\right\}\\
&=\max\left\{0, P_0\left(\sup_{f\in \mathcal{F}}\{\mathbb{G}^+(f) + c^+(f)\} > t_\alpha^{+}\right)+P_0\left(\inf_{f \in \mathcal{F}}\{\mathbb{G}^-(f) + c^-(f)\} < t_\alpha^{-}\right)-1\right\}.
\end{align*}
\end{proof}

\end{document}